\documentclass[useAMS,usenatbib]{mn2e}
\usepackage{times}
\usepackage{epsfig}
\usepackage{amssymb}

\def\be{\begin{equation}}
\def\ee{\end{equation}}
\def\bc{\begin{center}}
\def\ec{\end{center}}
\def\beq{\begin{eqnarray}}
\def\eeq{\end{eqnarray}}

\def\msun{{M_\odot}}

\def\gammac{\gamma_{\rm c}}

\def\gmax{\gamma_{\rm \max}}
\def\gmin{\gamma_{\rm \min}}
\def\Gammaf{\Gamma_{\rm f}}
\def\Gammaj{\Gamma_{\rm j}}

\def\doppler{\delta}

\def\phe{\epsilon}
\def\phx{x}
\def\xmin{\phx_{\min}}
\def\xmax{\phx_{\max}}

\def\sigmat{\sigma_{\rm T}}
\def\taut{\tau_{\rm p}}
\def\taugg{\tau_{\gamma\gamma}}

\def\taupb{t_{\rm pb}}

\def \Tmax{T_{\max}}
\def \Thetamax{\Theta_{\max}}

\def\d{{\rm d}}
\def\betaj{\beta_{\rm j}}
 
\def\etaiso{\eta_{\rm i}} 
\def\etab{\eta_{\rm B}} 
\def\etaj{\eta_{\rm j}} 
\def\etas{\eta_{\rm s}}

\def\etaeff{\varepsilon_{\rm eff}} 

\def\Lj{L_{\rm j}}
\def\Lk{L_{\rm k}}
\def\LB{L_{\rm B}}
\def\Ld{L_{\rm d}}

\def\Lph{L_{\rm ph}}
\def\Ldff{L_{\rm d,44}}

\def\Ldfv{L_{\rm d,45}}
\def\Ljfv{L_{\rm j,45}}

\def\ellj{\ell_{\rm j}}
\def\ellb{\ell_{\rm B}}
\def\elld{\ell_{\rm d}}
\def\elli{\ell_{\rm iso}} 

\def\Bj{B_{\rm j}}
\def\Bcr{B_{\rm cr}}

\def\RL{R_{\rm L}}
\def\Rg{R_{\rm S}}
\def\Rj{R_{\rm j}}

\def\Rpc{R_{\rm pc}}
\def\Rin{\varpi_{\rm in}}
\def\Rout{\varpi_{\rm out}}

\def\ergs{\rm erg\ s^{-1}}

\def\mpr{m_{\rm p}}
\def\me{m_{\rm e}}
\def\np{n_{\rm p}}

\title[High-energy emission of relativistic jets]
{Radiation from relativistic jets in blazars and the efficient 
dissipation of their bulk energy via photon breeding}

\author[Boris E. Stern and Juri Poutanen]
{Boris~E.~Stern$^{1,2,3}$\thanks{E-mail:
stern@bes.asc.rssi.ru (BES), juri.poutanen@oulu.fi (JP)}
and Juri~Poutanen$^{3}$\footnotemark[1] \\
$^{1}$Institute for Nuclear Research, Russian Academy of Sciences,
Prospekt 60-letiya Oktyabrya 7a, Moscow 117312, Russia\\
$^{2}$Astro Space Center, Lebedev Physical Institute,
Profsoyuznaya 84/32,  Moscow 117997, Russia\\
$^{3}$Astronomy Division, Department of Physical Sciences, 
P.O.Box 3000, FIN-90014 University of Oulu, Finland}

\begin{document}
\date{Accepted 2007 November 11. Received 2007 October 29; in original form 2007 September 19}
\pagerange{\pageref{firstpage}--\pageref{lastpage}} \pubyear{2007}
\maketitle

\label{firstpage}

\begin{abstract}
High-energy photons propagating in the magnetised medium with large velocity gradients
can mediate energy and momentum exchange. Conversion of these photons into electron-positron
pairs in the field of soft photons with the consequent isotropization and emission of new high-energy photons by Compton scattering can lead to the runaway cascade of the high-energy photons and electron-positron pairs fed by the bulk energy of the flow. This  is the essence of the  photon breeding mechanism.

We  study  the problem of high-energy emission of relativistic jets in blazars
via photon breeding mechanism  using 2D ballistic model for the jet with the detailed treatment 
of particle propagation and interactions. Our numerical simulations  from first principles 
demonstrate that a jet propagating in the soft radiation field of broad emission-line region
can convert a significant fraction (up to 80 per cent) of its total power into radiation. 

We show that  the gamma-ray background of similar energy density as 
observed at Earth  is sufficient to trigger  the photon breeding. 
The considered mechanism produces a population of high-energy leptons and,
therefore, alleviates the need for Fermi-type particle acceleration models 
in relativistic flows.
The  mechanism reproduces basic spectral features observed in blazars including
the blazar sequence (shift of spectral peaks towards lower energies with increasing luminosity).
The significant  deceleration of the jet at subparsec scales 
and the transversal gradient of the Lorentz factor (so called structured jet) predicted by the model 
reconcile the discrepancy between the high Doppler factors determined by the fits to
the spectra of TeV blazars and the low apparent velocities observed at VLBI scales.
The mechanism produces significantly broader angular distribution of radiation than
that predicted by a simple  model assuming the isotropic emission in the jet frame.
This helps to reconcile the observed statistics and luminosity ratio 
of FR I and BL Lac   objects with the large Lorentz factors of the jets as well  as to 
explain the high level of the TeV emission in the radio galaxy  M87.

We also discuss other possible sites of operation of the photon breeding mechanism and
demonstrate that the accretion disc radiation at the scale of about 100 Schwarzschild radii and 
the infrared radiation from the dust  at a parsec scale can serve as targets for photon breeding.
\end{abstract}

\section{Introduction}

The large gamma-ray isotropic luminosities  of the blazars  $10^{47}$--$10^{49}\ergs$ 
 \citep{muk97} require a very powerful energy source.   
There is a general consensus that this emission is produced by relativistic jets emanating  from 
the vicinity of a black hole. The beaming of radiation into a solid angle $\Omega\sim0.01$ sr
reduces the total luminosity to $10^{45}$--$10^{46}$$\ergs$, which is still comparable to the 
bolometric luminosity of a  luminous quasar. 
This implies a high efficiency  of the bulk energy conversion into  gamma-ray emission, 
because the accretion luminosity is similar to the jet power \citep{rs91}.

What kind of energy is available for the high-energy radiation of blazars:
the jet internal energy or the total energy carried by the jet
through the external environment? The first scenario can be associated with the
internal shocks \citep[e.g.][]{r78,spa01} or magnetic reconnection \citep[see e.g.][]{lut03}.
Then the efficiency is limited by the dispersion of the jet Lorentz factor or by the share of magnetic field
which can reconnect, which depends on the field structure.

The second scenario provides a larger potential energy budget, but requires
a ``friction'' between the jet and the external environment. A  promising
mechanism for such a friction can be associated with the exchange of neutral particles
between the two media which move with respect to each other. The simplest version of
such scenario is the exchange of photons by  Compton scattering \citep{ab92}.
If the energy flux in the quasar jet is dominated by cold protons then the Thomson optical
depth across the jet (assuming one electron per proton) is  
\be \label{eq:taut}
\taut(R)=  \frac{\Lj}{\Rj}  \frac{\sigmat}{2 \pi \mpr c^3 \Gammaj} \approx
2.3\times 10^{-5} \frac{\Ljfv}{R_{17} \theta \Gammaj}, 
\ee
where $\Lj$ is the jet power,  $\Rj$ is the radius of the jet cross-section, 
$R$ is the distance from the black hole, $\theta=\Rj/R$ 
and $\Gammaj$ are the jet opening angle and the initial Lorentz factor.\footnote{We use standard notations 
$Q=10^x Q_x$ in cgs units and for dimensionless variables.}
Such a small  optical depth is insufficient to provide a considerable photon viscosity.
However, the optical depth for high-energy photons against pair production on the
low-energy photons can be high enough. For example, a luminous accretion disc 
of a quasar emitting soft radiation with luminosity $\Ld \sim 10^{45}\ergs$ 
at a typical temperature $\Theta = kT/\me c^2= 10^{-5}$, provides a high 
opacity across the jet \citep[see ][ hereafter SP06]{sp06}: 
\be \label{eq:taugg}
\taugg(R) = 60 \frac{\Ldfv}{R_{17} \Theta_{-5}} (10\theta) ,
\ee
which remains more than unity up to the parsec scale.

\citet{der03} and \citet{ste03} suggested a mechanism of dissipation
of the bulk energy of relativistic fluids through the particle exchange with charge conversion (the converter mechanism). 
When this mechanism supplies energy to $e^{\pm}$ pairs and photons, it can work in a runaway manner similar to a chain 
reaction in a supercritical nuclear pile. Here the role of breeding neutrons belongs to the high-energy photons, 
which  extract energy from the fluid and breed exponentially. 
In this way the converter  mechanism takes the form of a supercritical photon breeding (hereafter, photon breeding).
\citet{ste03}  studied numerically the operation of the photon breeding mechanism 
in ultrarelativistic shocks propagating in a moderately dense environment,
showing the high efficiency of the bulk energy dissipation.
A similar problem for relativistic jets in active galactic nuclei (AGNs) was investigated by us (SP06), 
using a simple 1D ballistic model of the jet and a detailed treatment of particle
propagation and interactions. We showed that up to 20 per cent
of the jet kinetic energy can be converted into gamma-rays, if the following 
conditions are met: Lorentz factor above $\sim 4$, a sharp transition layer
between the jet and the external environment, the presence of the ambient  soft
photon field,  and a transversal or a chaotic, not very strong, magnetic field.

The present paper is a development of the approach presented in SP06. 
The main improvement is 2D treatment of the fluid (instead of 1D in SP06) and a better numerical resolution.
The 2D treatment allows us to track the evolution of the system for
much longer time, when it reaches a steady state. We also study the jet emission at
a wider range of distances and with different sources of the soft photon
background. We still neglect the internal pressure and treat the jet dynamics in a 
ballistic approximation.

In Section \ref{sec:numer} we describe our model of the jet, the external soft radiation
and the scheme of the numerical simulation. 
In Section \ref{sec:results} we describe the parameters of our simulations and 
present the results: the light curves of escaping photons, their spectra, distributions of the 
Lorentz factor of the decelerating jet, and the energy distribution of pairs produced in the jet.
The properties of the model and its limitations are discussed in Section \ref{sec:proper}. 
In Section \ref{sec:astro} we discuss the astrophysical implications of our model. 
We conclude in Section \ref{sec:concl}.

\section{Problem setup}
\label{sec:numer}

\subsection{Photon breeding mechanism}
\label{sec:pb}

Let us first summarise the basic physics of the photon breeding mechanism  in the context of 
relativistic jets (see SP06 for more details). 
Consider a jet with bulk Lorentz factor $\Gammaj$ propagating through  the stationary medium. 
Assume that the medium is filled with the background soft photons [e.g. from the accretion disc, 
broad-emission line region (BLR), dusty torus, etc.] 
and some seed high-energy photons (e.g. extragalactic gamma-ray background).  
An external high-energy photon of energy $\phe$ enters the jet and interacts with a soft (disc) photon producing an 
electron-positron pair.\footnote{The photon energies, in units of 
the electron rest mass $\me c^2$, are denoted as 
$\phx$ and $\phe$ for photons of low ($<1$) and high ( $>1$) energies, respectively.
We define also  the power-law index $\alpha \equiv - \d \log F(\phx)/\d\log \phx$, where $F(\phx)$ is the 
photon energy flux. } 
The typical energy of the high-energy photon interacting with the blackbody radiation field of temperature
$\Theta$ is a few times $\Theta^{-1}$.
The time-averaged Lorentz factor of the produced pair (as measured in the external frame) 
gyrating in the  magnetic field of the jet, becomes $\gamma\sim \Gammaj^2 \phe\sim \Gammaj^2\Theta^{-1}$. 
The electrons and positrons in the jet Comptonize soft photons (internal synchrotron or external)  
up to high energies.\footnote{Depending on the 
parameters the seed soft photons are either from the external medium or the synchrotron photons 
produced in the jet. In the first case, the inverse Compton  (IC) scattered radiation is called the
external radiation Compton (ERC) and in the second case, the synchrotron self-Compton (SSC). } 
Some of these photons leave the jet and produce pairs in the external environment, which
gyrate in the magnetic field and Comptonize soft photons more or less
isotropically. Some of these Comptonized high-energy photons enter the jet again.

In this cycle, the energy gain, $\sim$$\Gammaj^2$, is provided by the isotropization of the charged 
particles in the jet frame and is taken from the bulk energy of the flow. 
Other steps in the cycle are the energy sinks.  
The whole process proceeds in a runaway regime, with 
the total energy in photons and relativistic particles increasing exponentially, if 
the amplification coefficient (energy gain in one cycle) is larger than unity. 
As shown in  SP06, the amplification coefficient  is $A\lesssim \Gammaj^2/10$, implying the minimum 
Lorentz factor of the flow needed for the process to operate  $\Gammaj\sim$4.  
This theoretical minimum is, however, difficult to reach as other conditions should be met. 

The preferred conditions are: the weak jet magnetic field (which reduces the energy sink 
 by synchrotron radiation) and large energy density of the external isotropic radiation in the jet 
 frame (which increases the Compton losses). The latter requirement 
 translates into high $\Gammaj$ and/or high fraction of the disc radiation 
 scattered/reprocessed  in the ambient medium.
 A sufficiently sharp boundary between the jet and the external environment 
 also increases the efficiency of the process; however, as we show in this paper, it is 
 not a necessary condition.

\subsection{Model of the jet}

We consider a jet propagating from the centre of the quasar accretion disc of luminosity $\Ld$ 
through the   medium filled with the soft photons from the disc as well as scattered in the environment. 
The total jet power scales with the disc luminosity $\Lj=\etaj\Ld$.
The comoving value of the magnetic field in the jet $\Bj$ (its direction is
transversal by assumption) is related to the Poynting flux carried by the (two-sided) jet
\be\label{eq:LB}
\LB = \etab \Lj=  \frac{\Bj^2}{ 8 \pi} 2 \pi \Rj^2 \Gammaj^2 c \sim 8\times 10^{43}
\Bj^2 R_{17}^2 (\theta\Gammaj)^2 \ \mbox{erg\ s}^{-1} ,
\ee
which is assumed to be independent of distance (i.e. the magnetic field scales as $\Bj \propto 1/R$).
The ratio of the Poynting flux to the total jet power $\etab$ 
 affects the ratio of synchrotron and Compton losses.
The form of the rest of the energy 
content is not important in our approach as we ignore the internal pressure. 
We just assume that the jet does not contain high-energy interacting 
particles except those that are produced by the studied mechanism. 
If the rest of the jet energy is dominated by cold protons,  the jet  kinetic power 
(including the rest mass) is 
\be 
\Lk=  (1-\etab) \Lj= \Gammaj\ \dot{M}c^2 = 2\pi (R \theta)^2 \Gammaj \mpr c^3 \np ,
\ee
where $\dot{M}$ is the mass outflow rate and $\np$ is the proton concentration (in lab frame).
The total jet power is then $\Lj=\Lk+\LB$. 
 
\subsection{External environment}

\subsubsection{Accretion disc  radiation} 
\label{sec:accdisc}

As discussed in SP06, the photon breeding mechanism requires the existence of the 
external soft radiation field which is opaque for high-energy photons against pair production. 
The emission of the accretion disc is opaque for high-energy ($\sim 10^5$MeV) 
photons moving across the jet up to the distance of a few parsec for the disc luminosity $\Ld \sim 10^{45} \ergs$. 
We characterise  this radiation by the multicolour  disc spectrum assuming a  power-law 
$T(\varpi)= \Tmax (\varpi/\Rin)^{-3/4}$ dependence of temperature on the disc radius $\varpi$,
with the ratio of the outer to inner disc radius $\Rout/\Rin=10^4$. 
Here $\Tmax = 5$ eV is the maximum disc temperature at $\Rin=3\Rg$ and 
$\Rg=2GM/c^2$ is the black hole Schwarzschild radius. 
The dimensionless maximal temperature is $\Thetamax\equiv k\Tmax/ \me c^2 = 10^{-5}$. 
We neglect the expected dependence on the luminosity 
$\Tmax\propto \Ld^{1/4}$ as it is a small effect. 

At distances much larger than the disc size, the disc radiation is 
directed along the jet and does not interact with the gamma-rays produced in the jet. 
Thus the conversion of these gamma-rays into electron-positron pairs requires a 
source of transversal soft photons. Below we  consider three possibilities for such 
sources at different distance scales.

 A typical AGN in addition to the thermal UV radiation emits nonthermal
X-ray continuum extending up to $\sim 100$ keV constituting  about 10 per cent of
total AGN luminosity. This radiation was implemented in the model by SP06,
however,  in the broad emission line region (BLR) the X-rays provide very little
opacity for pair production, and therefore we do not include them in this work in 
order to reduce the number of parameters. 
Nevertheless, the X-rays can be important at smaller distances
from the black hole and their role is a matter of future studies.

\subsubsection{External radiation} 
\label{sec:softback}

\begin{itemize} 
\item Model A:  
 This model represents the isotropic radiation from the  BLR 
at a subparsec distance $R=2\times 10^{17}$ cm (as  in SP06). 
The soft external radiation in this case is the disc radiation 
reprocessed and scattered by clouds in the BLR with an admixture of a softer
infrared (IR) radiation from larger distances. 
The resulting spectrum  has a complicated form,
which we  parametrize here  (as in SP06) by a cutoff power-law 
$F_{\rm BLR} (\phx) \propto \phx^{-\alpha} \exp(-\phx/\xmax)$ 
extending from the far-infrared $\xmin \sim10^{-9}$ to the UV band, as we assume 
$\xmax= \Thetamax$.
The normalization of this spectrum is given by  the parameter $\etaiso\equiv U_{\rm iso}/U_{\rm d}$,
 which defines the ratio of the  energy density of the isotropic component to that of the direct disc radiation. 
We use $\etaiso= 0.05$  in most of the simulations.

The results are sensitive to the spectral slope $\alpha$ which describes the relative
number of soft photons. 
Here we set $\alpha = 0.4$ which  implies that the ratio of the isotropic energy density in the 
IR range to that of the direct disc radiation is $\sim 0.003$  (for $\etaiso=0.05$).
Such IR contribution can be supplied by the dust at a parsec scale which 
can absorb and reemit in the IR range a substantial part of the quasar luminosity.

\item Model B:  It describes the radiation field at a few parsec scale,  
where the isotropic component is dominated by the IR radiation of the dusty torus 
heated by the disc radiation. 
The isotropic component of the radiation field is represented as a blackbody
with temperature $T=600$ K ($\Theta = 10^{-7}$)  and $\etaiso=0.1$ or 0.3. 

\item Model C: No separate isotropic component  is present.  
The radiation field is provided by the accretion disc only  \citep{ds93}.
At $R\sim10^{16}$ cm  the angular size of the accretion disc is large enough to 
illuminate the jet under large angles. 
We take a standard multicolour disc, but for simplicity  assume that there is one-to-one 
correspondence between the emitted photon energy $\phx$ and the angle $\phi$ 
between the jet direction and the photon  momentum: 
\be
\tan \phi = \frac{\varpi}{R} = \frac{\Rin}{R} \left( \frac{\phx}{\Thetamax}\right) ^{-4/3}.
\label {eq:disc}
\ee
\end{itemize}

\subsubsection{External matter} 

The external environment  is kept at rest in spite of the fact that 
it gets a considerable momentum from photons producing pairs outside the jet. 
This momentum will cause an entrainment of the
external medium which is not easy to simulate. It is important that this 
motion should be nonrelativistic, otherwise the gradient of the Lorentz 
factor would be significantly lower than that in the simulations. We discuss the 
quantitative  condition for this in Section \ref{sec:external}. 

The magnetic field in the external medium $B_{\rm ext}$ is assumed to be 
purely transversal with its direction randomly distributed in the plane normal to the jet. 
The Larmor radius of the highest energy pairs, 
$R_{\rm L,16}=0.015 \gamma_8 /B_{\rm ext,-3}$,  should not exceed the jet radius, 
otherwise the efficiency of the breeding cycle is reduced. 
Therefore we take $B_{\rm ext}=10^{-3}$ G, 
which is two to three orders of magnitude higher than that in the interstellar medium, but still
realistic for the sub-parsec region of an AGN.

\subsection{Model parameters}

The efficiency of the considered mechanism depends  not directly on the luminosities,
but on the dimensionless compactness parameters. 
The compactnesses of the radiation and of the magnetic field can be defined as 
\be \label {eq:compac}
\ell = \frac{ \sigmat }{\me c^2} U \Rj,
\ee
where $U$ is the corresponding energy density. The compactness is useful 
when estimating the cooling rate of pairs. 
The comoving magnetic compactness is defined as
\beq\label{eq:ellb}
\ellb & = & \frac{ \sigmat }{\me c^2} \frac{\Bj^2}{8\pi}  \Rj = 
3.2\times 10^{-4} \Bj^2 R_{\rm j,16}  \nonumber \\
&= &   
4.3\times 10^{-3} L_{\rm B,45}  R_{\rm j,16}^{-1} \left( \frac{\Gammaj}{10}\right)^{-2}.
\eeq
The photon compactness, defined in the external frame, has  
two components: the direct accretion disc radiation
%(see details in Section \ref{sec:accdisc}, \ref{sec:softback})
\be
\elld = \frac{\sigmat}{\me c^2} \frac{2\Ld}{4\pi R^2 c} \Rj = 
4.3\times 10^{-3} \Ldfv  R_{17}^{-2} R_{\rm j,16} 
\ee 
(with factor 2 in front of $\Ld$ coming from the Lambert law) 
and the isotropic  component  $\elli=\etaiso \elld$. 
With these definitions, we can write the equation for electron energy losses  
in the jet frame (for model A)
%(primes correspond  to the values measured in the jet frame) 
as 
\be \label{eq:cooling2}
\frac {\d\gamma}{\d t} \approx  - ( \ellb +   \elli \Gammaj^2 + \elld /\Gammaj^2) \gamma^2 ,
\ee
where the term with $\ellb$ describes the synchrotron losses and the second and third terms -- 
the Compton losses (in Thomson limit) on the isotropic radiation field
(with a factor $\sim\Gammaj^2$ coming from the Lorentz transformation
to the jet frame) and on the direct disc radiation, respectively. 
Because the soft photons can interact with high-energy pairs in
Klein-Nishina regime, the second term in reality is much reduced. 
%The disc radiation can be neglected as its energy density  in the jet frame  
%is reduced by a factor $\sim\Gammaj^2$. 
Equation (\ref{eq:cooling2}) does not include  SSC losses, which are zero 
at the beginning of the photon breeding cascade 
(because it takes time to built up the synchrotron energy density),  
but, in principle, can be significant at the steady-state regime. 
The pairs produced outside of the jet lose their energy mostly by Compton scattering of 
the disc radiation (the  synchrotron losses are negligible because of 
the low magnetic field there) with $\elld$ determining the cooling rate:
\be \label {eq:cooling_ext}
\frac {\d\gamma}{\d t} \lesssim  -  \elld \gamma^2 .
\ee 

The jet compactness  
\be
\ellj = \frac{\sigmat}{\me c^2} \frac{\Lj}{2\pi \Rj c} = 
0.43\  L_{\rm j,45}  R_{\rm j,16}^{-1} 
\ee 
is related to the Thomson optical depth through the   jet 
\be
\taut = \frac{\me}{\mpr}  \frac{\ellj}{\Gammaj}  (1-\etab) = 
2.3 \times 10^{-5}  (1-\etab)   L_{\rm j,45}  R_{\rm j,16}^{-1}  \Gamma_{\rm j,1}^{-1}  .
\ee

\subsection{Simulation method}

\subsubsection{Radiative processes} 

The  numerical simulation of radiative processes is based on the Large Particle
Monte-Carlo code (LPMC) described in  \citet{st85} and \citet{sbss95}. 
The present version of the code handles Compton scattering, synchrotron radiation,
photon-photon pair production, and pair annihilation. All these processes are
reproduced without any simplifications at the micro-physics level.
Synchrotron self-absorption was neglected as it consumes too much computing power 
and is not important  at the considered conditions. 
The number of large particles (LPs) is $2^{19} = 524288$ for most 
of the simulations (in SP06 we used $2^{17}$ LPs), which 
is still insufficient to completely get rid of numerical 
fluctuations. A substantially better resolution will require  massive parallel computations
and a smoothing technique which has to be developed. 
 
\subsubsection{Particle kinematics}

The scheme of particle tracking in the relativistic fluid was developed in \citet{ste03} and 
described in details in  SP06. 
The 4-momentum of electron/positron LPs are defined in the comoving frame
of the fluid. 
If the electron/positron has a critical Lorentz factor  
\be \label{eq:gmax}
\gmax  \approx  10^8 {\Bj^{-1/2}} ,
\ee 
it loses half of its energy in a half of Larmor orbit  to synchrotron radiation. 
This requires a fine particle tracking at the first Larmor orbit for  the comoving 
Lorentz factor of the particle $\gamma \gtrsim \gmax$, and,  
therefore, the comoving tracking step is limited to $\Delta t = 0.1 \RL/c$, 
where $\RL = 1.7 \times 10^3 \gamma/\Bj\ \mbox{cm}$ is the Larmor radius. 

In the external medium, synchrotron losses are small and an electron 
is cooled by the disc radiation. The condition that the electron cooling time-scale 
is longer than the inverse of Larmor frequency can be written as 
\be
\frac{1}{\gamma\elld} > \frac{2\pi \RL}{\Rj} ,
\ee 
which translates to the limit on the Lorentz factor 
\be
\gamma< \gamma_{\rm ext, max} =  10^6 
\left(  \frac{ R_{\rm j,16}  B_{\rm ext,-3} }{\ell_{\rm d,-3}} \right)^{1/2} . 
\ee

In our simulations, we exactly track  the particle motion 
for the duration of the first Larmor orbit for the comoving $\gamma> 2\times 10^5$. 
For smaller $\gamma$ and at further orbits we neglect 
the dependence between the comoving direction and the energy and 
sample the direction of the particle assuming a uniform distribution 
of its gyration phase  in the comoving system (see details in SP06). 

The trajectories of electrons and positrons in the magnetic field are
simulated assuming a random direction of the field in the plane
normal to the direction of motion 
% transversal geometry 
in the jet as well as in the external matter. The magnetic field correlation 
length is assumed to be larger than the Larmor radius.

\subsubsection{Geometry and jet dynamics} 

\begin{figure}
\centerline{\epsfig{file=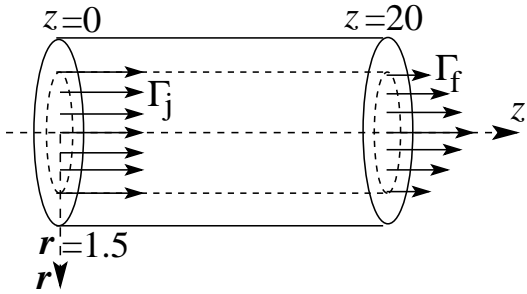,angle=270,width=7.0cm}}
\caption{
Schematic representation of the jet geometry. 
The jet is represented by a cylinder of radius $r=1$. The simulation volume of radius 
$r=1.5$ includes also the external medium.  At the inlet, $z=0$, the jet has Lorentz 
factor $\Gammaj$. At the outlet, $z=20$, the terminal Lorentz factor $\Gammaf$ depends 
on radius $r$. 
 }
\label{fig:geom}
\end{figure}

Our model of the jet is the same as in SP06 except now we use 2D representation
of the fluid instead of the 1D. 
We consider a piece of the jet  centred at distance $R$ from the central 
source and length 20 $\Rj$, where the jet radius  is $\Rj=R \theta$ and 
$\theta$ has the meaning of the opening angle, while 
for simplicity we approximate the jet  by a cylinder of radius $\Rj$.
The boundary condition at the inlet 
is the flow of constant Lorentz factor $\Gammaj$ (see Fig.~\ref{fig:geom}).\footnote{Our model 
assumes a well-collimated jet with $\theta\lesssim 1/\Gammaj$. 
If the opening angle of the jet is much larger, 
then the efficiency of photon breeding drops, because 
the probability for the high-energy photons  to escape  the jet decreases.} 

We introduce dimensionless coordinates, where the transversal distance from the jet axis 
$r$ and the distance from the central source $z$ are measured in units of  $\Rj$.  
We use  a fixed Eulerian grid with fairly rough spacing: 20 cells in the $z$-direction 
and 64 cells in the $r$-direction. 
The step on $r$ is nonuniform:  the spacing is finer near the jet boundary.
The time unit is $\Rj/c$. 

The fluid in each cell $i$ is characterised by its total 
energy $E_i$ and momentum $P_i$, which define the cell Lorentz factor:   
\begin{equation}
\Gamma_i = {E_i\over \sqrt{E_i^2 - P_i^2} } . 
\end{equation}
The flow is treated as one-dimensional  along the cylinder axis, 
because the transversal  component of $P_i^r$ never 
exceeds $10^{-2}$ of the total momentum and the transversal displacement of the flow 
can be neglected.

\subsubsection{Statistical representation} 

For a good statistical representation of the process, it is necessary 
to have more or less constant density of the LPs over the volume of the cylinder. 
As the energy (and number) density of particles (pairs and photons) grows 
by orders of magnitude from the inlet of the cylinder 
to the outlet, we need to introduce the dependence of statistical weights  on $z$.
The statistical weight $w$ of a particle of energy $\phe$ is defined through 
the energy content  $w \phe =  2^n\ s$, where $n$ is the integer part of $z/2$ 
for $z <14$ and $n=7$ at $z > 14$.
 The value $s$ varies with time in order to keep the total number of LPs 
approximately constant. 
This weighting scheme is implemented as follows: we split 
a particle into two identical particles with twice less statistical 
weight when it crosses one of the planes $z = 2, 4, 6 ...$ in the direction opposite 
to the jet motion and we kill a particle with probability 0.5 and increase its weight 
by 2,  when it crosses the plane in the direction of jet motion. The price for 
a good resolution at the region, where the avalanche starts and grows, is 
a poor resolution at the output. For that reason the Poisson noise 
in output light curves and photon spectra is rather high.

\subsubsection{Simulation setup}

At the start of simulations we set constant Lorentz factor $\Gammaj$ 
over the jet volume ($r<1$ and  $0<z<20$, see Fig.~\ref{fig:geom}) and 
assume a sharp boundary between the jet and
the surrounding medium.  
We also inject a number of high-energy photons with the 
total energy several orders of magnitude smaller than the steady-state luminosity of the jet. 
(There is no injection of any additional photons later during the simulations and 
the high-energy population of particles is self-supporting.)
The external soft-photon radiation field is constant during the simulation. 

Interactions between particles change the energy and momentum of the fluid. 
The electron-positron pairs born in the process become part of the fluid.
The momentum and energy exchange between photons and each grid cell of the flow 
is accumulated during the time step $\Delta t = 0.01 \Rj/c$ and the total 
4-momentum of the cell is updated after that. 
The momentum is transferred between cells along the jet axis.

The output of the simulations consists of the photons that  escape 
from the cylindric ``simulation volume'' with the boundaries $r = 1.5$, $z=0$, and $z = 20$. 
We also record the energy spectrum of photons and pairs in the simulation volume.

\section{Results}

\label{sec:results}

\subsection{Simulation runs}

%The parameters of the simulations are chosen to correspond to the quasars. 
We consider the distance scale of $R \sim 10^{16}$--$10^{19}$ cm and 
the jet Lorentz factor $\Gammaj\sim 8$--$40$. 
We vary the luminosity of the accretion disc $\Ld$  in the interval $\sim 10^{43}$--$10^{46} \ergs$ 
and try different ratios of the jet to the disc power.
First, we present results of three series of simulations, where we vary
one of the parameters and nine non-systematic runs covering a wider area of 
parameters to demonstrate the working area of the mechanism. 
In the two series we vary the disc and jet luminosity, while in the third series we 
vary the jet Lorentz factor. 
Most of our simulation runs concern model A. 
Three runs correspond to different astrophysical scenarios  described by models B and C, 
where the photon breeding operates.  

%\renewcommand{\theenumi}{(\alph{enumi})}
%\renewcommand{\theenumi}{(\arabic{enumi})}
%\renewcommand{\labelenumi}{\theenumi}

%\begin{enumerate}
\begin{enumerate}
\item  Runs 01--05: This series represents magnetically dominated jet (i.e. $\etab=1$) with 
the jet power equal to the disc luminosity ($\etaj=1$), 
which is varied between $\Ld =  3 \times 10^{45}\ergs$ down to $3 \times 10^{43}\ergs$,
where the photon breeding becomes sub-critical and the process does not work 
(at  $R=2\times 10^{17}$). 

\item  Runs 10--16:  Similar to the previous series, but for the matter dominated jet with $\etab=0.2$ and
$\etaj=1$. We vary $\Ld$ in a wider range, because the process for smaller $\etab$ has a lower luminosity threshold. 

\item Runs 21--25: In this series we have fixed the disc luminosity to $\Ld = 3 \times 10^{44}\ergs$ 
use $\etaj=1$ and $\etab=0.2$,  and vary the initial Lorentz factor of the jet. 
 
\item Runs 31-35: Run 30 demonstrates a photon breeding process starting from the extragalactic gamma-ray background
and resulting in 20 orders of magnitude growth of the high-energy photon population
(see Section \ref{sec:highseed}).
Runs 31 represents a strongly non-linear system, where the jet power exceeds the disc luminosity.
Run 32 is the case of the weakest AGN ($\Ld = 2 \times 10^{43}\ergs$) tried in this work for model A.
Run 33 demonstrates that a high efficiency can be achieved at $\Gammaj = 10$ if the magnetic field is weak.
Run 34 is a trial with the lowest Lorentz factor, $\Gammaj = 8$, and small Poynting flux.
Run 35 is similar to run 12, but for the harder spectrum of isotropic external photons 
($\alpha = 0$ instead of  $\alpha = 0.4$). 

\item Runs 41--42: They correspond to Model B, where we simulate interaction of a strong jet 
$\Lj = 2.3\times10^{45}\ergs$ of Lorentz factor $\Gammaj=20$ with the IR radiation 
($\Theta = 10^{-7}$) of the dusty torus at a parsec scale.  
Run 43 describes Model C, where a relatively weak,  $\etaj = 0.1$, jet 
interacts with the direct radiation from the accretion disc of $\Ld = 8\times 10^{43}\ergs$
at a distance of a hundred gravitational radii (for a $10^8$ solar mass black hole).

\end{enumerate}

The parameters for these runs are summarised in Table 1.

\begin{figure}
\centerline{\epsfig{file=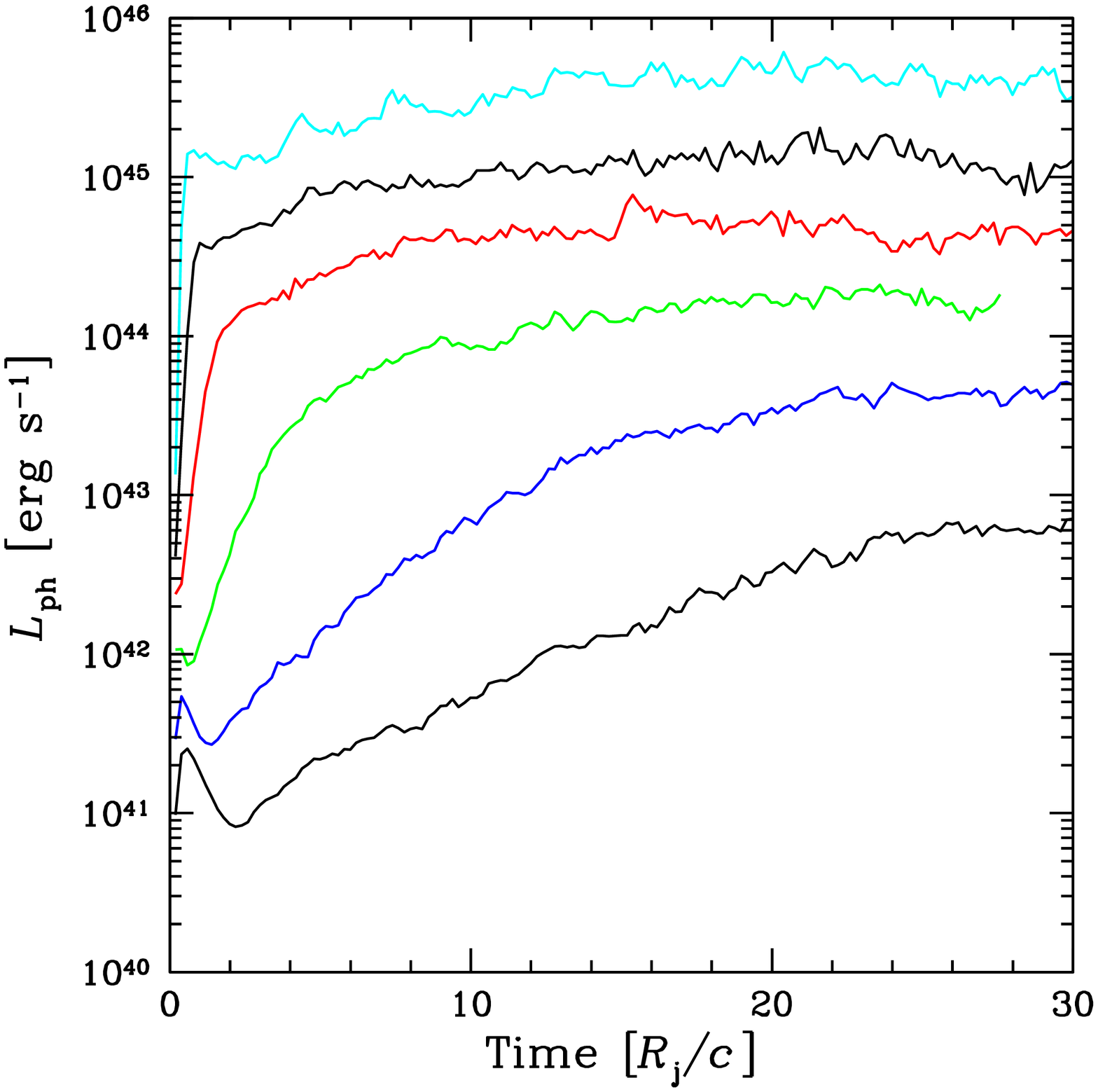,width=7.5cm}}
\caption{The power  converted into photons (including those born in the jet, but not yet escaped) 
versus time for series of runs 10--15 (from top to bottom).   }
\label{fig:evol}
\end{figure}

 \begin{figure}
\centerline{\epsfig{file=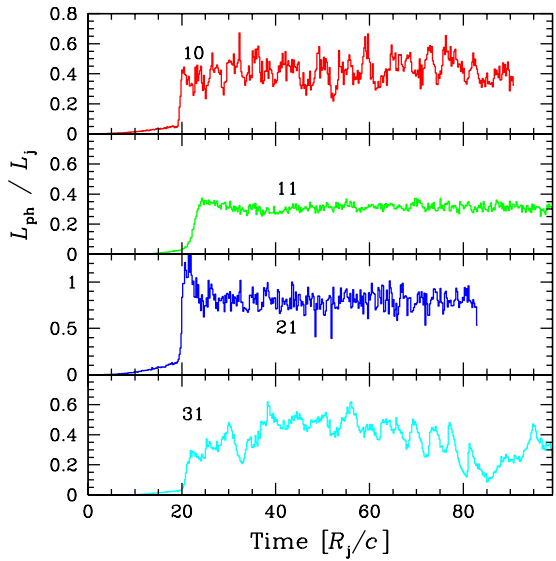,width=7.5cm}}
\caption{The escaping photon luminosity in units of the jet power versus time for runs 10, 11, 21 and 31.
Note that the light curves look different compared to Fig.~\ref{fig:evol} as it takes 
about 20 time units for photons to escape from the simulation cylinder (and the ordinate scale is linear here). 
}
\label{fig:licurv}
\end{figure}

%%%%%%%%%%%%%%%%%%%%%%%%%%%%%%%%%%%%%%
\begin{table*}
\caption{Model parameters and results.}
 \centering   
 \begin{minipage}{175mm}
  \begin{tabular}{@{}llrlr llrrr rrlclrl@{}}
  \hline
Run 		&  $R$$^a$ & $\Gammaj$$^b$  & $\theta$$^c$& $\Ld$$^d$ & ER$^e$ & $\etaiso$$^f$  & $\etaj$$^g$  &  $\etab$$^h$ 
&$\elld$$^i$  & $\Gammaj^2 \elli$$^j$ &$\ellb$$^k$ & $B$$^l$ &  $\taut$$^m$ &  $\taupb$$^n$ & $\Gammaf$$^o$  & $\etaeff$$^p$ \\ 
  &  cm &   &  & $\ergs$  &  &    &  &   &   &  & &  G  &  & & &  \\ 
  \hline
01		&2E17&20&0.05& 3E45 &A& 0.05 & 1.0 & 1.0 & 3.3E-3 & 6.6E-2 & 3E-3 & 3.2   & 0 & 0.5 &18.0 & 0.25 \\
02 		&2E17&20&0.05& 1E45 &A& 0.05 & 1.0 & 1.0 & 1.1E-3 & 2.2E-2 & 1E-3 & 1.8   & 0 & 1.1 &18.0  & 0.32  \\
03 		&2E17&20&0.05& 3E44 &A& 0.05 & 1.0 & 1.0 & 3.3E-4 & 6.6E-3 & 3E-4 & 1.1   & 0 & 3.7 &18.6 & 0.26  \\
04		&2E17&20&0.05& 1E44 &A& 0.05 & 1.0 & 1.0 & 1.1E-4 & 2.2E-3 & 1E-4 & 0.6   & 0 & 6.2 &19.6 & 0.05  \\
05		&2E17&20&0.05& 3E43 &A& 0.05 & 1.0 & 1.0 & 3.3E-5 & 6.6E-4 & 3E-5 & 0.35 & 0 & --   & 20.0 & 0    \\
& \\ %  \hline 
10		&2E17&20&0.05& 1E46 &A& 0.05 & 1.0 & 0.2 & 1.1E-2 & 2.2E-1 & 2E-3 & 2.6   & 9.3E-5 & 0.1  &16.2& 0.42  \\
11		&2E17&20&0.05& 3E45 &A& 0.05 & 1.0 & 0.2 & 3.3E-3 & 6.6E-2 & 6E-4 & 1.4   & 2.8E-5 & 0.2  &16.2& 0.44  \\
12		&2E17&20&0.05& 1E45 &A& 0.05 & 1.0 & 0.2 & 1.1E-3 & 2.2E-2 & 2E-4 & 0.82 & 9.3E-6 & 0.42 &14.7& 0.50 \\
13		&2E17&20&0.05& 3E44 &A& 0.05 & 1.0 & 0.2 & 3.3E-4 & 6.6E-3 & 6E-5 & 0.45 & 2.8E-6 & 1.2  &13.6& 0.56  \\
14		&2E17&20&0.05& 1E44 &A& 0.05 & 1.0 & 0.2 & 1.1E-4 & 2.2E-3 & 2E-5 & 0.26 & 9.3E-7 & 3.0  &15.6& 0.41  \\
15		&2E17&20&0.05& 5E43 &A& 0.05 & 1.0 & 0.2 & 5.5E-5 & 1.1E-3 & 1E-5 & 0.18 & 4.6E-7 & 5.2  &18.6& 0.12  \\
16		&2E17&20&0.05& 3E43 &A& 0.05 & 1.0 & 0.2 & 3.3E-5 & 6.6E-4 & 6E-6 & 0.14 & 2.8E-7 & --     &20.0& 0        \\
& \\ %  \hline 
21		&2E17&40&0.05& 3E44 &A& 0.05 & 1.0 & 0.2 & 3.3E-4 &2.6E-3 & 1.5E-5 & 0.22 & 1.4E-6  &  0.62 &10.0 & 0.82   \\
22		&2E17&30&0.05& 3E44 &A& 0.05 & 1.0 & 0.2 & 3.3E-4 &1.5E-3 & 2.7E-5 & 0.30 & 1.8E-6  & 0.78  &10.3 & 0.77   \\
23=13 	&2E17&20&0.05& 3E44 &A& 0.05 & 1.0 & 0.2 & 3.3E-4 &6.6E-3 & 6E-5    & 0.45 & 2.8E-6  & 1.2     &13.6& 0.56   \\  
24		&2E17&14&0.05& 3E44 &A& 0.05 & 1.0 & 0.2 & 3.3E-4 &3.2E-3 & 1.5E-4 & 0.71 & 3.9E-6  & 3.0     &13.6& 0.14    \\
25		&2E17&12&0.05& 3E44 &A& 0.05 & 1.0 & 0.2 & 3.3E-4 &2.4E-3 & 1.9E-4 & 0.80 & 4.6E-6  & --       &12.0&  0      \\
& \\ %  \hline 
30		&2E17&20&0.05& 1E44  &A& 0.05 & 1.0 & 0.12  & 1E-4 & 2E-3   & 1.3E-5 &   0.2   & 1.0E-6  & 2.7  &12.7& 0.53 \\
31	 	&2E17&20&0.05& 1E45  &A& 0.05 & 9.0 & 0.13   & 1E-3 & 2E-2   & 1.3E-3 &  2      & 9.0E-6   & 0.95 &17.7& 0.30  \\
32 		&2E17&20&0.05& 2E43  &A& 0.05 & 1.0 & 5E-3 & 2.3E-5 & 4E-4 & 1.3E-7 & 0.02 & 2.3E-8   & 7.0  &19.5& 0.03  \\
33 		&2E17&10&0.05& 3E44  &A& 0.05 & 0.8 & 1.3E-2 & 3E-4 &1.5E-3 & 1.3E-5 & 0.2 & 5.9E-6    & 1.8  &8.7& 0.31  \\
34 		&2E17& 8 &0.05& 3E44  &A& 0.05 & 1.0  & 7E-3 & 3E-4 &1E-3 &1.3E-5  &  0.2      & 9.3E-6    & 2.6  &7.7& 0.13 \\
35$\approx$12 &2E17&20&0.05&1E45&A$^q$&0.05&1.0&0.2&1.1E-3&2.2E-2& 2E-4  & 0.82  & 9.3E-6  &1.8  &16.2& 0.41\\
& \\ %  \hline 
41 		&6E18&20&0.033& 1.3E45  &B& 0.3 & 1.8 & 2E-2 & 3E-5 & 3.7E-3 & 2.6E-6  & 0.02  & 1.3E-6  & 5.4  &18.0& 0.21  \\
42 	 	&6E18&20&0.033& 1.3E45  &B& 0.1 & 1.8 & 6E-3 & 3E-5 & 1.2E-3 & 6.4E-7  & 0.01  & 1.3E-6   & 6.3   &19.1& 0.14  \\
43	 	&6E15&20&0.067&    8E43  &C&  0   & 1.0 & 6E-2 & 4E-3 &    0      & 1.3E-5    & 1.0    & 2.2E-6   & 2.3  &19.2& 0.19  \\
\hline
\end{tabular}
\label{tab:runs}
{ $^{a}$Distance from the black hole.
$^{b}$Initial Lorentz factor of the jet.  
$^{c}$Half-opening angle of the jet. 
$^{d}$Total disc luminosity. 
$^e$Model of the external radiation described in Section \ref{sec:softback}.
$^{f}$Ratio of the isotropic soft photon energy density to 
that of the disc.  
$^{g}$Ratio of the total jet power to the disc luminosity, $\Lj/\Ld$.  
$^h$Fraction  of the  jet power carried by the Poynting flux, $\LB/\Lj$.  
$^i$Disk compactness.
$^j$Compactness of the isotropic external radiation in the jet frame. 
$^k$Magnetic compactness. 
$^l$Magnetic field in the jet. 
$^m$Thomson optical depth through the jet produced by electrons associated with protons.
$^n$e-folding time of the photon breeding  in $\Rj/c$ units. 
$^o$Terminal  Lorentz factor at the jet axis. 
$^p$Average  efficiency of the jet energy conversion into radiation in the steady-state. 
%$^o$Number of LPs in the simulations was $2^{21}$. 
$^q$Modified model A with a harder spectrum of external isotropic photons ($\alpha = 0$). 
}
\end{minipage}
\end{table*}
%%%%%%%%%%%%%%%%%%%%%%%%%%%%%%%%%%%%%%

\subsection{Time evolution and the steady-state regime}  

The energy release curves for set 2 (runs 10--15) are presented in Fig.~\ref{fig:evol}. 
All curves show an exponential rise at the start with the time 
constant depending on the disc luminosity. 
The fast  exponential rise  turns to a slower growth when approaching the steady state.  
A very fast rise (time constant $\sim 0.1 \Rj/c$  for run 10) 
occurs in the case of a dense external radiation field when the photon free path 
length is much smaller than the jet radius. Then the photon breeding takes place only in a narrow 
layer at the jet boundary. Such fast breeding terminates after a short time, when the thin boundary 
layer decelerates. The breeding turns to a slower mode, where the softer photons with a longer 
free path dominate.

Note that the injection of external high-energy photons takes place only 
at the start of the simulation. Then the regime becomes self-supporting.
The reason for this is a spatial feedback loop. 
The photon avalanche  develops from a smaller to a large $z$, but  some high-energy 
photons produced in the external environment move in the opposite direction, 
from larger $z$, where there are many high energy photons, to 
smaller $z$, where they can initiate a new photon avalanche. 
Due to this feedback a high efficiency of emission $\etaeff = \Lph/\Lj$ can be reached 
even  if the growth of the photon avalanche is slow.

We can see large fluctuations of the energy release at the saturation regime
(at the stage of the exponential rise, the curves are smooth). Fluctuations 
have the Poisson component which is large because of the weighting method:
LPs at the outlet have statistical weight  $2^7$ larger than at the inlet.
There are also non-Poisson fluctuation which are
 caused by fluctuations in the spatial feedback: a photon  with a large statistical 
weight moves from the end  of the active jet fragment to its beginning, interacts there and 
produces a photon avalanche. 
The number of such photons is small, but  the amplification coefficient 
(i.e. the number of secondary photons and pairs)  can exceed $10^4$.  
Therefore the number of LPs as large as $\sim 10^6$ is still insufficient for a good statistical representation.

However, statistical noise is not the only   reason for the 
large fluctuations in the output photon flux. In runs 10 and 31,  where the compactness
is high, and especially in run 31, where the system is strongly non-linear,  we see a 
large amplitude variability at a range of time-scales (see Fig.~\ref{fig:licurv}).
We performed run 31 with the different number of LPs: $2^{19}$ and $2^{21}$, but 
the character and the amplitude of the fluctuations was similar. 
Thus,  in this case we deal with the real instability of a non-linear dynamical system
 (see Section \ref{sec:variability} for discussion).

\subsection{Radiative efficiency and the ``working area'' of the process}

\label{sec:radeff}

The resulting efficiency of the conversion of jet power into radiation at the steady-state 
and the time constant of the exponential photon breeding are shown in the last columns of Table1. 
Once we fixed $R$, $\Rj$, and the ratio $\etaiso$  in all runs for model A, 
our main remaining parameters are the jet Lorentz factor, the disc luminosity and the Poynting flux. 
The jet power is not important unless
it exceeds the disc luminosity (as in run 31),  making the system strongly non-linear.
The series of runs 01--05 shows that the process can work efficiently in the case of
a magnetically dominated jet giving $\etaeff\approx1/4$. 
The lowest  luminosity when it works  is $\sim 10^{44}\ergs$.
If the matter dominates over the magnetic energy by a factor of 5 (runs 10--16)
then the radiative efficiency is higher, above 50 per cent, 
and the working luminosity range slightly extends down to $5 \times 10^{43}\ergs$. 
The supercritical breeding is still possible at even lower disc luminosity, $2 \times 10^{43}\ergs$ (see run 32),
but only if the magnetic field is very weak. 
Below this threshold the photon-photon opacity 
across the jet (see equation \ref{eq:taugg}) becomes insufficient.
(We note here that the working luminosity limits are given for 
$R = 2 \times 10^{17}$ cm, $\Gammaj=20$ and $\etaiso=0.05$.)

Remarkably, the efficiency at a moderate disc luminosity ($\sim3\times 10^{44}\ergs$, run 13) is 
higher than for a more luminous quasar (runs 10, 11 and 31). 
The reason is that at a high luminosity the spine of the jet is shielded 
from the high-energy external photons by the opaque (respectively to pair production) 
radiation of the accretion disc and synchrotron radiation of pairs in the jet. 
In runs 10 and 31, only the photons with energy $\phe < 10^3$ 
can penetrate deeply into the jet. Nevertheless, such moderate energy 
photons still can  interact with the synchrotron photons (mostly produced in the decelerated outer 
jet layers) inside the jet and 
support the photon breeding.

 %######################################
\begin{figure*}
\begin{center}
\leavevmode \epsfxsize=7.0cm \epsfbox{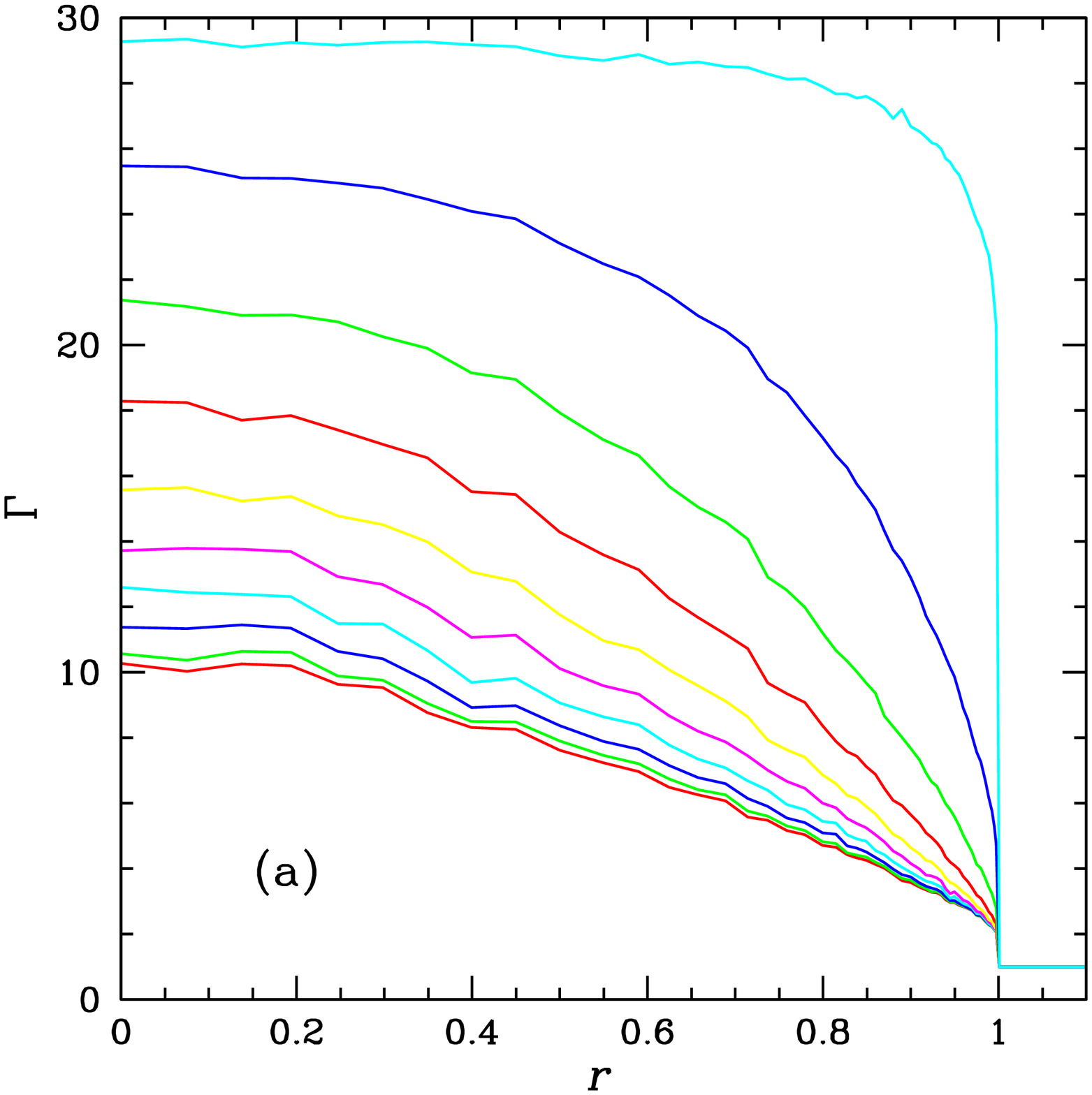} \hspace{1cm}
\epsfxsize=7.0cm \epsfbox{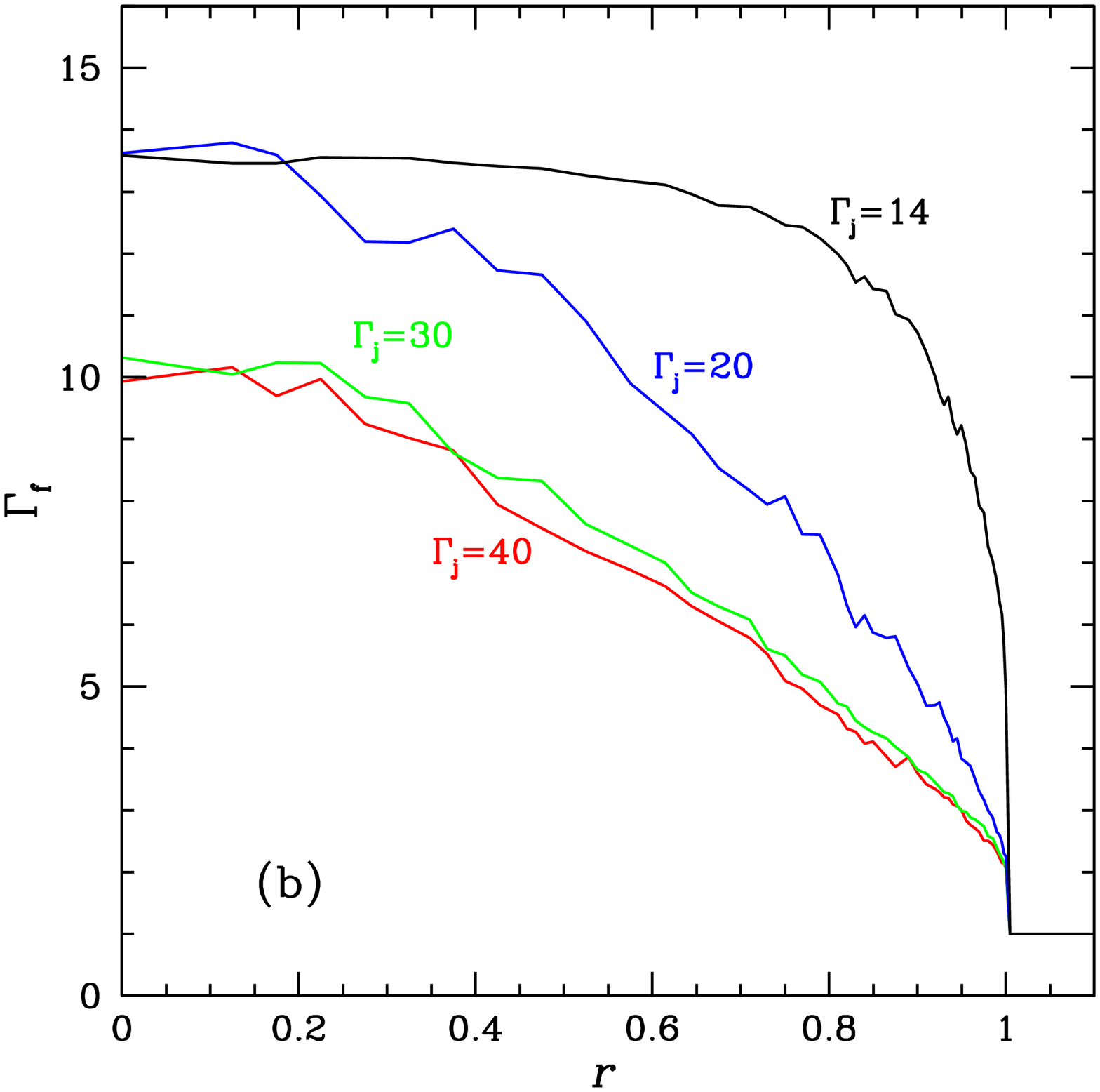}
\end{center}
\caption{ \label{fig:lorlev}  
The transversal distribution of the bulk Lorentz factor.  
(a) Dependence on distance $z$ for run 22 
(curves from top to bottom correspond to $z=2$ with step 2 till  $z=20$). 
(b) Distribution of the terminal Lorentz factor at the outlet ($z=20$) for runs 21--24 
corresponding to $\Gammaj=$40--14. }
\end{figure*}
%######################################

 As one can expect, the dependence on $\Gammaj$ is very strong. 
The efficiency $\etaeff$ is close to $(\Gammaj -\overline{ \Gammaf })/\Gammaj$, where 
the average Lorentz factor at the outlet $\overline{\Gammaf}$ slightly 
{\it decreases} when  $\Gammaj$ increases (see Section \ref{sec:jet_dec}).

At a lower $\Gammaj$ the process becomes very sensitive to the ratio of the 
Poynting flux to the isotropic disc luminosity. Indeed, at the given magnetic flux
$\LB$ the comoving magnetic compactness scales as $\ellb \propto \LB/\Gammaj^2$ 
(see Eqs. \ref{eq:ellb}), while Compton cooling rate 
in the Thomson limit scales as $\propto \elli \Gammaj^2$, see equation (\ref{eq:cooling2}).
Thus the ratio of Compton to synchrotron losses scales as $\etaiso\Gammaj^4 \theta^2/\etab$. 
%(see e.g. SP06). 
%\be
%\frac{U'_{\rm iso} }{U'_{\rm B}} = \frac{\etaiso}{\etab}  \Gammaj^4 \theta^2.
 %\ee 
This ratio decreases somewhat because of the Klein-Nishina effect, but 
 it is still a strong function on $\Gammaj$. 
That is the reason why  the runaway photon breeding does not happen at 
$\Gammaj <14$ if $\etab = 0.2$ (series 21--25). 
If the magnetic field is very weak, then the process can work efficiently at 
$\Gammaj = 10$ (run 33) and less efficiently, 
but still in  a self-supporting regime at  $\Gammaj = 8$ (run 34). 

All results described above are obtained under an assumption that the BLR 
is relatively abundant with isotropic IR photons, so that the photon 
spectrum can be roughly approximated as 
a cutoff powerlaw  with $\alpha = 0.4$ (see Section~\ref{sec:softback}). 
The spectrum however can be harder if, for example,  there is no sufficient amount of dust 
around the AGN. Therefore, we made a couple of trials for model A with a harder BLR spectrum,  $\alpha = 0$.
One trials  with the same parameters as run~13 
($\Ld = \Lj = 3 \times 10^{44}\ergs$) resulted in no self-supporting
photon breeding. Another trial with ($\Ld = \Lj = 10^{45}\ergs$ (run 35  in Table~1)  
demonstrated photon breeding,  
which was  slower than in run 12 (with $\alpha=0.4$). However, the final steady state was
almost the same including the radiative efficiency (0.41 versus 0.5 in run 12) and the emitted spectrum. 
Such a striking difference between the two cases differing only by factor 3 
in luminosity appears because of  the non-linear character of the mechanism for high-power jets:
soft synchrotron photons produced by the jet significantly contribute to the photon-photon opacity. 
The problem with soft photon starvation can be resolved  if some
internal jet activity (e.g. internal shocks) supply soft synchrotron photons which can be then 
scattered in the BLR \citep{gm96}. 

In general, the results of our simulations for distance $R$ from the black hole 
and luminosity $L$,   can be  scaled to another distance $R'$ if 
one changes at the same time all luminosities as $L' =L R'/R $. 
In this case, the compactnesses of all radiation components and
of the magnetic energy do not change (see equation \ref{eq:compac}). 
Such scaling is not exact,  because the value of magnetic field  
changes and the synchrotron spectra change accordingly. 
The scaling, however, does not affect the criticality of the system as 
long as the ratio between synchrotron and Compton losses remains constant. 

At luminosity smaller than about $L_{\rm d,\min} \approx 10^{43} R_{17} \ergs$ 
the opacity for pair-production through the jet becomes smaller than unity  and the 
super-criticality condition breaks down. 
Assuming the distance to the emission region of $R \sim 10^3 \Rg$, we can 
express the luminosity threshold in terms of the Eddington luminosity 
$L_{\rm d,\min} \sim 10^{-5} L_{\rm Edd}$.

The distance to the BLR may also depend on the luminosity of the object. 
The reverberation mapping gives the relation 
$R_{\rm BLR,17} \approx \Ldff^{1/2}$ \citep{pet93,kas05}. 
This then gives the condition $\Ldff >  0.1 \Ldff^{1/2}$, which translates to the 
lower limit on the luminosity $\Ld > 10^{42}\ergs$, when the photon breeding can operate.

The conditions when model A operates can be summarised as follows:
\begin{itemize}
\item At $\Gammaj\gtrsim 20$  and at $L_{\rm d,44} \sim L_{\rm j,44} \gtrsim 5 R_{17} $ the photon breeding 
mechanism is efficient and very robust: it works independently on the share of magnetic energy in the jet and
is not sensitive to the spectrum of the external radiation.

\item At $\Gammaj\gtrsim 20$ and lower jet power  $L_{\rm j,44}  < 5  R_{17} $,
 when the disc luminosity is still sufficiently high, $L_{\rm d,44} >  0.5 R_{17}$, 
the mechanism is sensitive to the spectrum of isotropic radiation in the BLR and, if there is a deficit 
of soft photons, the photon breeding is not self-supporting.

\item For $\Gammaj <20 $ the mechanism becomes sensitive to the magnetic field in the jet. 
The strong field inhibits the photon breeding. When the magnetic energy is much smaller that 
the matter kinetic energy, the efficient emission is possible down to $\Gammaj \approx 8$.

 \item The threshold of the disc luminosity 
% (assuming the fraction $\etaiso=0.05$  being reprocessed in the BLR)
is $L_{\rm d,44} \sim 0.1 R_{17}$. 
However, the self-supporting emission at such low $\Ld$ requires weak
magnetic field and a soft spectrum of the  BLR radiation. 
\end{itemize}

We have also tried models B and C  just to show that 
the process can work in different situations. 
Radiative efficiencies of 20 per cent are easily reached. 
We do not perform a systematic study of 
these models because both scenarios depend on various conditions that require a different study. 
Model C, for example, should be considered together with the mechanism of 
the jet launching and in model B one should account for the history of the 
jet flow before it reaches the parsec-scale distance.

%######################################
\begin{figure}
\centerline{\epsfig{file=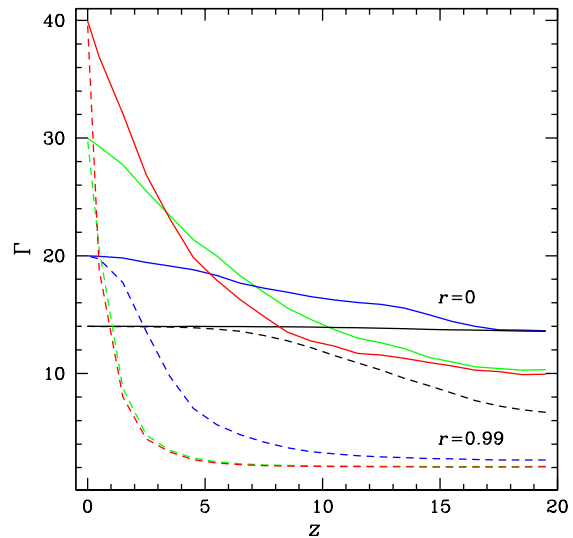,width=7.5cm}}
\caption{Distribution of the fluid Lorentz factor in $z$ direction for runs 21--24. 
Solid curves show the Lorentz factor at the jet axis $r=0$, and the 
dashed curves are for the jet boundary at $r=0.99$. 
}
\label{fig:gamz}
\end{figure}
 %######################################

%######################################
\begin{figure*}
\begin{center}
\leavevmode \epsfxsize=7.0cm \epsfbox{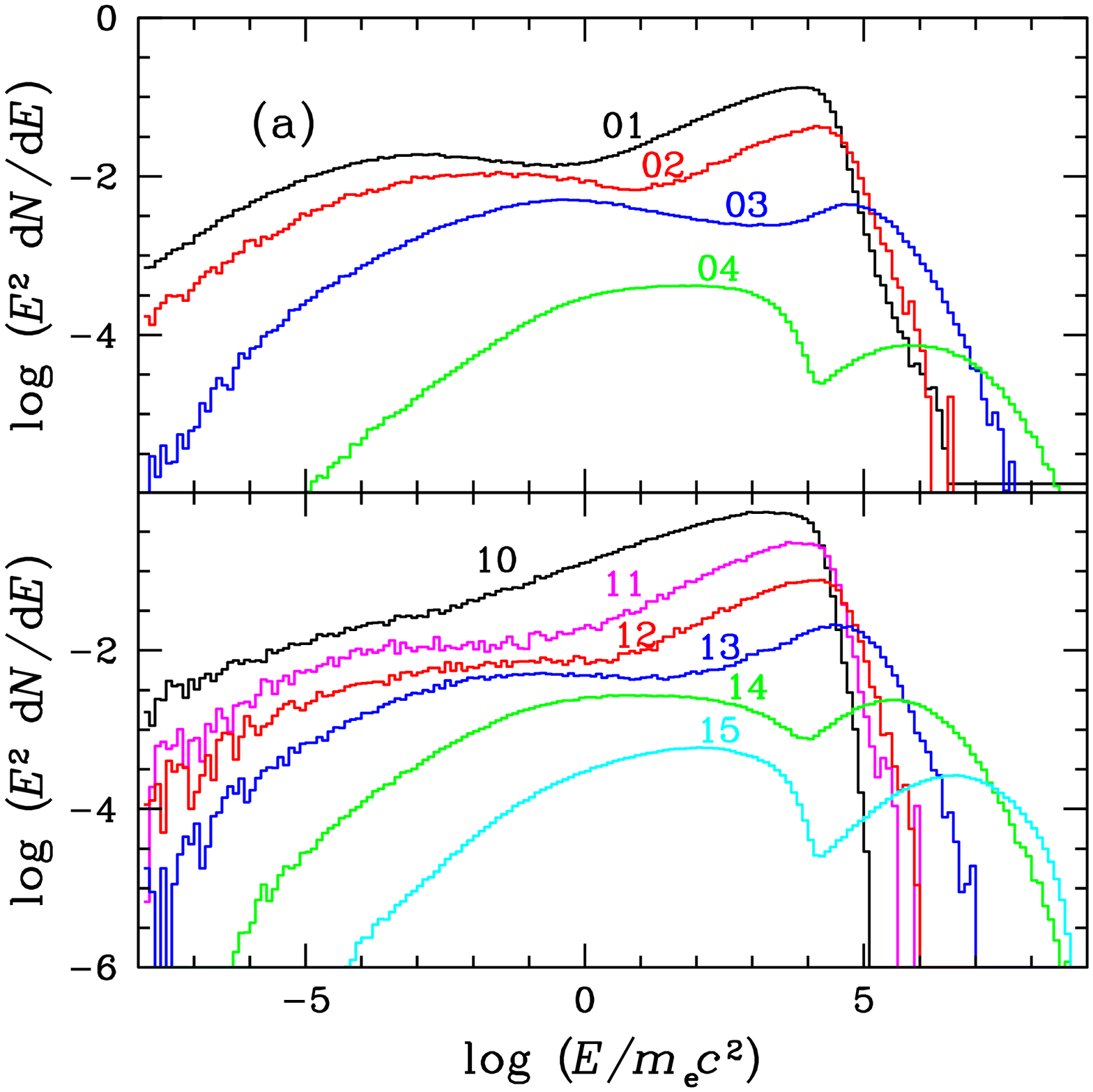} \hspace{1cm}
\epsfxsize=7.0cm \epsfbox{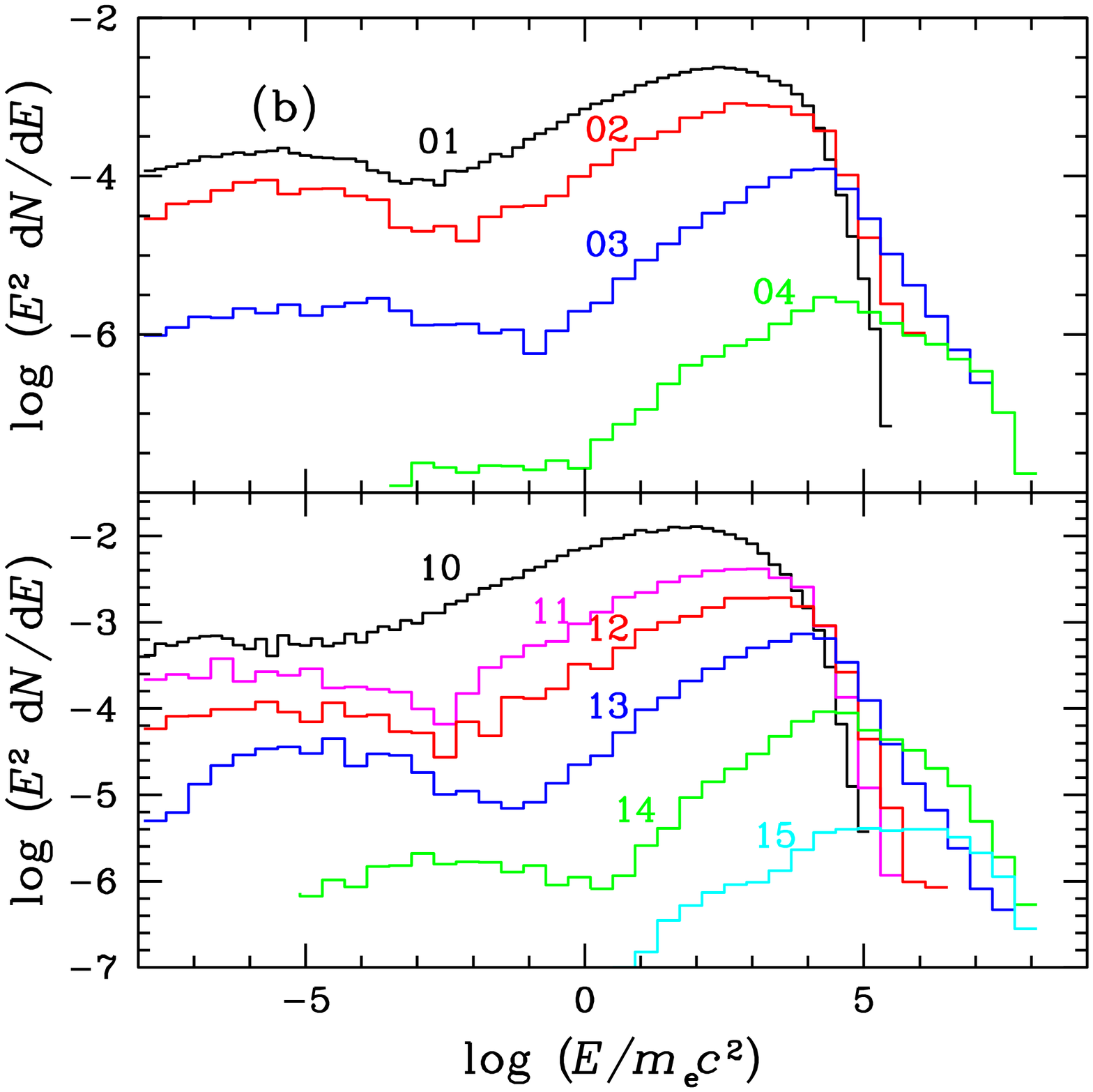}
\end{center}
\caption{ \label{fig:spec}  
Spectra of the escaping photons (angle-integrated) for runs 01--04 (upper panel) and 
runs 11--15 (lower panel).
(a) Radiation spectra emitted along the jet axis with $\theta<10\degr$. 
(b) Radiation spectra at large angles to the jet axis, $\theta>40\degr$.
}
\end{figure*}
%######################################

\subsection{Jet deceleration and the energy release}
\label{sec:jet_dec}

Jet deceleration in the 1D approach of  SP06 took place mainly near the
jet boundary ($r >0.9$), while the centre of the jet was not affected.
In the 2D simulations we can track the deceleration of the deeper parts of the jet up to its  axis. 
The transversal distributions of the fluid Lorentz factor for runs 21--24 are   
shown in Fig.~ \ref{fig:lorlev}. 
Distributions of $\Gamma$ along the jet axis are shown in Fig.~\ref{fig:gamz}.
It is remarkable that the terminal Lorentz factor $\Gammaf$ depends weakly on the 
initial $\Gammaj$, actually it even  decreases when  $\Gammaj$ increases.

The highest gradient of the Lorentz factor is near the jet boundary. 
 In  run 22, the Lorentz factor 
near the boundary drops from 30 down to 5 at $z = 3$ (see Fig.~\ref{fig:gamz}). 
The jet experiences  similar deceleration 
in runs 01, 10, 11 and 31, where the radiation density is high.
At the outlet the gradient of $\Gammaf$ depends weakly on the distance from the axis.  
 It reaches some critical value beyond which the photon breeding is not effective anymore. 
In cases of a lower compactness (runs 04, 14, 15, 32) the Lorentz factor changes 
more smoothly with $z$. 
At the outlet,  $\d\Gamma/\d  r$ is still  large close to the boundary, with the 
jet getting the form of  a fast spine and a slow sheath. 
A large gradient of $\Gamma$ means that our ballistic approximation rapidly breaks down   
in the cases of a large compactness or a large initial Lorentz factor. 
We discuss the consequences   in Section \ref{sec:ballis}.

\subsection{Radiation spectra}

Figure \ref{fig:spec}(a) shows the output spectra of photons emitted in the direction of the jet ($\theta < 10\degr$). 
The spectra (except those for runs 10 and 31 with very high compactness) 
show a two-component synchrotron--Compton  structure.
Runs 01--03 produce a synchrotron peak in the range $\phe \sim 10^{-3}$--$10^{-1}$, consistent with those 
observed in  BL Lacs, however, the theoretical peaks are less prominent that the observed ones
(detailed comparison with the observed spectra is presented in Section \ref{sec:sychmax}), 
partially due to averaging over angles. 
The Compton component  at high $\Ld$ and $\Lj$ (runs 01, 02, 10--12, 31) has a 
cutoff at $10^4<\phe< 10^5$ due to the photon-photon absorption: 
the external isotropic BLR radiation provides a larger than unity optical depth for the high-energy photons.

The spectra for a smaller compactness (runs 04, 14, 15, 32, as well as runs 41, 42, where the 
compactness is small because of large $R$), have prominent synchrotron peaks at 
high energies $\phe \sim 10^{0}$--$10^{3}$. 
The synchrotron and Compton components are clearly separated above  
the maximal energy of synchrotron radiation: 
\be \label{eq:vemax}
\phe_{\rm s, \max} \sim \Gammaj \gmax ^2 \Bj/\Bcr \approx 230 \Gammaj,
\ee
which does not depend on the magnetic field (here $\Bcr=4.4\times 10^{13}$ G is the critical  field). 
The value of $\gmax$ given by equation (\ref{eq:gmax}) 
is defined as the maximal electron energy (in the jet frame)   
after half the Larmor orbit, i.e. when it turns around from the counter-jet direction to the 
jet propagation direction (see SP06). 
The derived $\phe_{\rm s, \max}$ assumes that the magnetic field is uniform at the scale of
the electron Larmor radius and is transversal to the jet. 
It is not sensitive to the soft radiation field because an electron with the comoving 
Lorentz factor $\gamma  \gtrsim 10^{8}$
interacts with the external radiation in the deep Klein-Nishina regime and the synchrotron losses dominate. 
Thus, the model predicts a new spectral feature which, in principle, can be observed by {\it GLAST}. 
Its detection would be a clear signature of the photon breeding mechanism
(a distribution of diffusively accelerated electrons will hardly extend to 
the energy, where the electron loses a large fraction of its energy in one Larmor orbit).
Moreover, because of independence of $\phe_{\rm s, \max}$ of the magnetic 
field and other parameters,  the Lorentz factor of the jet can, in principle, be determined
from the detailed fits of the shape of the spectral cutoff.

For the considered model parameters, the dominant source 
of seed soft photon for Compton scattering is BLR.  
At the start of simulations, the SSC losses are small, because 
the synchrotron energy density needs to be built up from zero. 
In the steady-state regime, the ratio  of  the SSC to the 
synchrotron losses can be expressed as $\etaeff \eta_{\rm s, KN} /\etab$,
where  $\eta_{\rm s, KN}$ is the fraction of  {\it soft} synchrotron radiation 
(i.e. Comptonized in Thomson regime) in the total jet luminosity. 
We can estimate from Fig.~\ref{fig:spec}(a) that  $\eta_{\rm s, KN} < 0.3$ for runs 01--03 
(with $\etab = 1$) and  $\eta_{\rm s, KN} < 0.1$ for runs 10--12 (with $\etab = 0.2$). 
For a lower jet and disc luminosity (runs 03--04, 13--15), the electrons in the jet
have higher energies and the SSC losses are suppressed by the Klein-Nishina effect.
Therefore the SSC losses are always at least a few times smaller than the synchrotron
losses and usually are much smaller than the ERC losses.
However, the Compton losses of pairs in the fast spine of the jet on the
synchrotron photons from the slower sheath can be considerable,
at least in runs 10 and 31 with the highest compactness.
These cases can be considered as  intermediate  between SSC and ERC.

%######################################
\begin{figure}
\centerline{\epsfig{file= 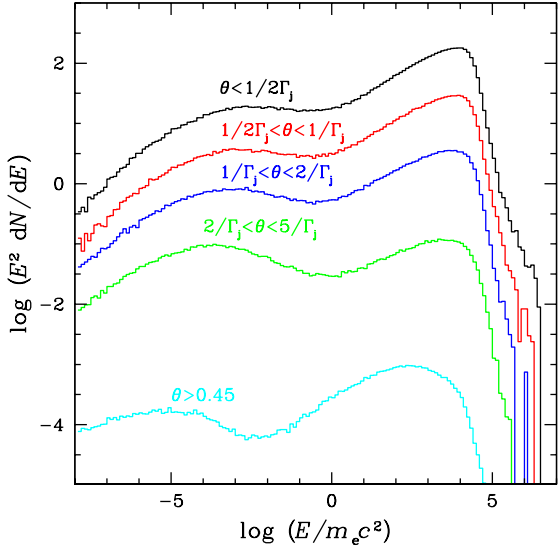,width=7.5cm}}
\caption{Intensity of radiation (per unit solid angle) 
at different angles to the jet axis for run 01.
}
\label{fig:spe_ang}
\end{figure}
%######################################

The radiation spectra  at large angles relative to the jet axis are shown in Fig.~\ref{fig:spec}(b).
We again see two-component, synchrotron  and Compton spectra.
The total (angle-integrated) luminosity in the gamma-rays emitted at large angles to the jet axis 
is about two order of magnitude smaller than that of the beamed emission,
which approximately corresponds to the gain factor $\Gammaj^2$.
 The radiation spectra emitted as a function of the angle $\theta$ 
between the line of sight and the jet axis are shown in Fig.~\ref{fig:spe_ang}. 
We see that the synchrotron peaks shift to lower energies at larger angles
and  the two components are better separated.

\subsection{Angular distributions of radiation} 
\label{sec:ang_dist} 

\subsubsection{Radiation from isotropic pairs in the steady jet}

\label{sec:ang_isojet} 
Let us first discuss what kind of angular distributions can be expected 
from a steady (narrow) axisymmetric jet in relativistic motion. 
The observed bolometric luminosity is related to the emitted one (in the jet frame) 
by the relation \citep[see e.g.][]{lb85,sik97}
\be
L(\theta) \equiv   \frac{\d L}{\d \Omega} = 
\frac{\doppler^3}{\Gammaj} \frac{\d L'}{\d \Omega'} 
\ee
where $\doppler=1/\Gammaj(1-\betaj\cos\theta)$ is the Doppler factor, 
and $\d \Omega=\d\phi \d\cos\theta$. 
Let us define  the amplification factor as the ratio of the observed and emitted luminosities
\be
A(\theta) \equiv \frac{L(\theta) }{L'/4\pi} . 
\ee
If the radiation source is isotropic in the jet frame, then $\d L'/\d \Omega' =L'/4\pi$, and 
\be \label{eq:isojet}
A(\theta) = \frac{\doppler^3}{\Gammaj} .
\ee
Integrating over angles, we can verify that the total emitted and 
observed powers are the same. Equation (\ref{eq:isojet})
can be applied, for example, to the synchrotron radiation in a tangled magnetic field 
or to  the SSC radiation by isotropic electrons. 

%######################################
 \begin{figure}
\centerline{\epsfig{file= 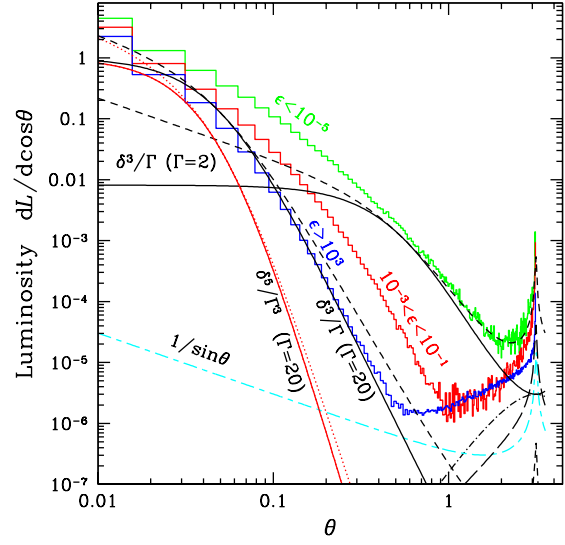,width=7.5cm}}
\caption{Luminosity (per unit solid angle) as a function of angle between the 
light of sight and the jet axis for run 01 in the three energy bands: 
IR-optical ($\phe<10^{-5}$), X-rays ($10^{-3}<\phe<10^{-1}$) and GeV gamma-rays ($\phe>10^{3}$)
(histograms, arbitrary normalization).  
The theoretical dependencies (\ref{eq:isojet}) expected for the steady relativistic jet emitting 
isotropically in the comoving frame  shown by solid (black) curves (cases with $\Gamma=2$ and 20). 
Dashed curves represent the synchrotron and SSC emission from  pairs in 2D circular 
motion, given by  equation (\ref{eq:isoph_2D_jet}).
The solid (red) curve describes the jet ERC radiation from isotropic pairs (equation \ref{eq:erciso}), 
while the dotted (red) curve is for the jet ERC radiation from   
2D pairs (equation \ref{eq:ERC_2D_jet}). 
The dot-dashed curve represents the emission of isotropic pairs 
in the external medium Comptonizing radiation beamed along the jet (equation \ref{eq:ICdisc}). 
The emission from 2D pairs in the external medium is shown
by long dashes (IC of beamed emission, see equation \ref{eq:2D_disc_ext}) and 
short-long dashes (synchrotron as well as IC of the isotropic photons, 
see equation \ref{eq:IC2D_isoph_ext}). 
}
\label{fig:angle}
\end{figure}
%######################################

If  soft photons are isotropic in the external medium, 
they are strongly beamed in the direction opposite to  jet propagation 
in the jet frame. 
This results in the  angular distribution of the radiation 
IC scattered by relativistic isotropic electrons \citep[see e.g.][]{rl79}
\be \label{eq:ICbeam}
\frac{\d L'}{\d \Omega'} = \frac{L'}{4\pi} \frac{3}{4} (1 +\cos \theta')^2 ,
\ee
with $\theta'$ being the angle to the direction of jet propagation. 
For small $\theta$, we can rewrite $1+\cos\theta'  \approx \doppler/\Gamma$, and 
we get a much sharper  angular dependence 
of the ERC radiation \citep[see also][ and Fig.~\ref{fig:angle}]{d95}: 
\be \label{eq:erciso}
A(\theta) =    \frac{3}{4} \frac{\doppler^5}{\Gammaj^3} .
\ee
Now the integration over all solid angles shows that the total observed power 
is 3/2 larger than the emitted one, 
which results from the non-zero net momentum of the 
emitted radiation  in the jet comoving frame (\ref{eq:ICbeam}).

\subsubsection{Radiation of primary pairs}

In the above we assumed explicitly an isotropic electron distribution. 
However, if the jet magnetic field 
has a  transversal geometry (as assumed in this paper), 
the distribution may significantly deviate from the isotropic one. 
For example, the first generation of pairs, produced in the jet
by the external high-energy photons (which are beamed in the jet frame similarly 
to the external soft photons),  has a very narrow distribution of pitch angles around  $\pi/2$.   
Let us now derive the angular distribution of radiation expected from these ``2D''  pairs. 

The pairs gyrating in a  transversal magnetic field
have a   flat distribution in $\theta$ (not $\cos\theta$). 
Therefore, their synchrotron emission has a pattern 
\be \label{eq:2Dsyn_ssc_cf}
 \frac{\d L'}{\d \cos \theta'}  = \frac{L'}{4\pi}  \frac{1}{\pi}   \frac{1}{\sin\theta'} .  
\ee
An identical distribution would be produced by SSC emission 
(because the synchrotron photons that are Comptonized by 2D pairs are mostly produced 
by the next generation, lower energy, isotropic  pairs). 
With  the relation $\sin\theta'=\doppler\sin\theta$, we get 
the amplification factor   (see dashed curves in Fig.~\ref{fig:angle})
\be \label{eq:isoph_2D_jet}
A (\theta)  =  \frac{1}{\pi} \frac{\doppler^2}{\Gammaj}  \frac{1}{\sin\theta} . 
\ee
This distribution has sharp peaks at $\theta=0$ and $\pi$. 

The external radiation IC scattered  on 2D pairs produces a pattern 
\be \label{eq:2Derc}
\frac{\d L'}{\d \cos \theta'}  = \frac{L'}{4\pi}  \frac{2}{3\pi}   \frac{(1 + \cos\theta')^2}{\sin\theta'} ,  
\ee
which gives the distribution of the observed emission  with 
the peaks at $\theta=0, \pi$ (see dotted curve in Fig.~\ref{fig:angle}): 
\be \label{eq:ERC_2D_jet}
A (\theta)  =  \frac{2}{3\pi} \frac{\doppler^4}{\Gammaj^3}  \frac{1}{\sin\theta} . 
\ee

\subsubsection{Radiation from external environment}

In addition to the radiation from the jet, there is also radiation 
from the relativistic pairs produced in the external environment. 
They also have two populations. The first population is 
produced by the high-energy photons 
from the jet propagating a small angles to the jet direction $\theta\lesssim1/\Gamma$. 
They perform 2D circular motion in the transversal magnetic field of the surrounding. 
The second population  produced by the pair cascade is nearly isotropic. 

The synchrotron radiation of 2D pairs follows distribution (\ref{eq:2Dsyn_ssc_cf}), 
with $\theta'$ replaced by $\theta$ (see short-long dashes in Fig.~\ref{fig:angle}). 
\be \label{eq:IC2D_isoph_ext}
L (\theta) \propto \frac{1}{\pi}   \frac{1}{\sin\theta} .  
\ee 
The IC scattered isotropic BLR photons have the same distribution. 

However, because most of the seed photons (synchrotron photons from the jet or 
the accretion disc photons) are beamed in the direction of the jet, the IC 
emission by the external pairs follows the distribution (\ref{eq:2Derc})  
with $\theta'$ replaced by $\pi-\theta$ (see long dashes in  Fig.~\ref{fig:angle}): 
\be \label{eq:2D_disc_ext}
L (\theta)  \propto   \frac{2}{3\pi}   \frac{(1 - \cos\theta)^2}{\sin\theta} .
\ee 
Thus, the 2D pairs produce a sharp peak at $\theta=\pi$.

The emission of isotropic pairs Comptonizing the disc (or jet synchrotron) radiation 
follows the law (\ref{eq:ICbeam}) with $\theta'$ replaced by $\pi-\theta$ 
(see dot-dashed curve in Fig.~\ref{fig:angle}): 
\be \label{eq:ICdisc}
L (\theta) \propto \frac{3}{4}  (1-\cos \theta)^2 . 
\ee
And, finally, isotropic pairs scattering isotropic radiation obviously 
produce a flat distribution of $L(\theta)$.

%######################################
\begin{figure}
\centerline{\epsfig{file=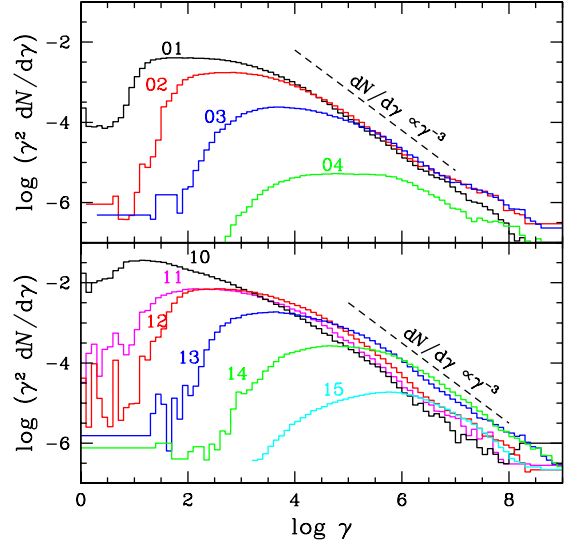,width=7.5cm}}
\caption{Electron (and positron) energy distributions in the jet comoving frame 
averaged over the jet volume for runs 01--04 and 10--15. }
\label{fig:espec}
\end{figure}
%######################################

\subsubsection{Emission pattern: simulations} 
\label{sec:angle}

As an example of the angular distribution of radiation predicted 
by the photon breeding model, we present results for run 01. 
The histograms in Fig.~\ref{fig:angle} show the dependence 
of the luminosity in three typical energy bands  (IR-optical, X-ray and gamma-rays).
We see that the distribution of high-energy photons at small $\theta$ 
follows quite closely that expected from the jet of $\Gammaj=20$ emitting 
isotropically in its rest frame (e.g. SSC, given by equation \ref{eq:isojet}).   
The ERC radiation from the jet is more beamed (see equation \ref{eq:erciso} and the
red solid curve in Fig.~\ref{fig:angle}). 
An excess  at $\theta=0$ can be explained by the contribution 
of the SSC and ERC emission from 2D pairs 
(given by eqs. \ref{eq:isoph_2D_jet} and \ref{eq:ERC_2D_jet}, respectively; 
see dashed and dotted curve in Fig.~\ref{fig:angle}).

At angles $\theta\gtrsim 0.45$, another component is obvious. 
This large-angle component is nearly isotropic and its flux at infinity is
4.5 orders of magnitude smaller than the average flux observed  within the jet opening.  
For the run 01 with $\Gammaj = 20$, the observed luminosity of this large-angle component  
at $\theta=\pi/2$ exceeds the luminosity given by expression (\ref{eq:isojet}) 
by more than 3 orders of magnitude. 
This off-axis emission is produced by the pairs gyrating in the external environment via 
Compton scattering of the  disc and BLR photons  and synchrotron photons from the jet. 
A peak at $\theta=\pi$ can be well described by IC emission 
of 2D pairs in the external environment (see long dashes and short-long dashes in
Fig.~\ref{fig:angle}) given by formulae (\ref{eq:IC2D_isoph_ext}) and (\ref{eq:2D_disc_ext}).

The low-energy (IR-optical  and the X-rays) photons are much less beamed.  
The spatial gradients of $\Gamma$ is the main cause of that. 
For example, the average $\Gamma$ of the inner half of the jet volume with $r<0.7$ in 
run 01 is about 18,  while the outer 20 per cent move with $\Gamma\lesssim10$, dropping 
to $\Gamma\sim 2$ in the outer 5 per cent (see also Figs~\ref{fig:lorlev} and \ref{fig:gamz}).

The soft band is dominated by the synchrotron radiation (Fig.~\ref{fig:spe_ang}). 
The emission at $\theta=0.5$ exceeds the simple estimate~(\ref{eq:isojet}) 
by 3 orders of magnitude. 
At angles $\theta\gtrsim 1/\Gammaj=0.05$ the emission is mostly produced in the slower, 
decelerated layers of the jet (see Fig.~\ref{fig:lorlev}), and the emission at 
$\theta\gtrsim 0.5$ is produced in the layers of $\Gamma\sim2$.  
The  synchrotron from 2D pairs in the jet even produces a peak at $\theta=\pi$. 
Some contribution to this peak is provided by the IC (low-energy tail) emission 
from the external medium.

The emission in the X-rays is the mixture of the synchrotron and ERC emission resulting 
in the intermediate behaviour seen in Fig.~\ref{fig:angle}. The large-angle emission here is 
produced by IC  in the external environment  (see Fig.~\ref{fig:spe_ang}) and 
therefore has an angular dependence similar to that of the gamma-rays. 

The sharp feature at $\theta=0$  will be washed out if the jet is actually conical rather 
than cylindrical or the jet magnetic field is tangled. 
The tangled magnetic field in the external medium will also smooth out 
the predicted peak at $\theta=\pi$.

% the jet Lorentz factor 
%varies across the simulation volume approximately as 
%\be
%\Gamma(r,z)= 1+ \left( \Gamma_0 -1 - \frac{z}{8}\right) \left( 1- r ^\xi\right)^{1/2}, 
%\ee
%where $\Gamma_0=20$ is the initial Lorentz factor at the inlet and $\xi\approx 3 +15/z$. 
%
% L(\theta) \propto \int_V  \doppler^5/\Gamma^3 \frac{\d \Gamma(r,z)}{\d r} \d^3 {\bf r} 
%
%

\subsection{Pairs produced in the jet}
\label{sec:pairs}

%######################################
\begin{figure}
\centerline{\epsfig{file=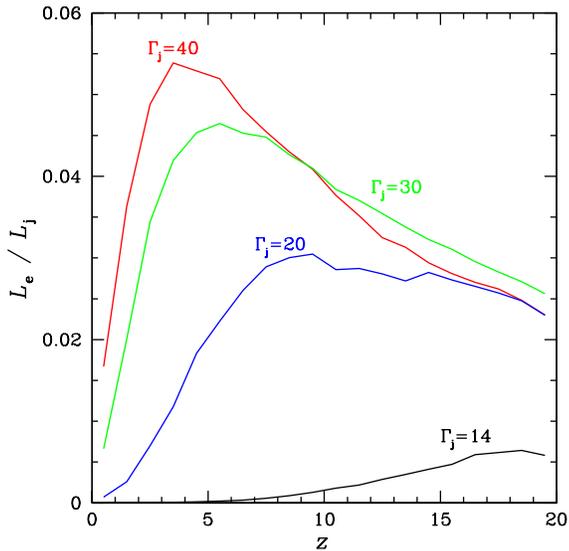,width=7.5cm}}
\caption{Kinetic luminosity carried by pairs with respect to the initial jet power
for runs 21--24 ($\Gammaj=$40--14).  }
\label{fig:elz}
\end{figure}
%######################################

The electron and positron comoving energy distributions averaged over time and jet volume 
for runs 01--04 and 10--15 are shown in Fig.~\ref{fig:espec}. 
All spectra except that for run 15 have a clear low-energy cutoff  at $\gammac=10$--10$^3$, this is the 
energy the pairs  cool down to  in the dynamical time. 
The cutoff comoving energy assuming the Thomson regime for Compton cooling 
and neglecting SSC can be expressed as 
\begin{equation}
\gammac = \frac{1}{t_{\rm dyn} (\ellb + \elli \Gamma^2)},
\end{equation}
where $t_{\rm dyn}\approx 10/\Gamma \sim 1$ is a typical (comoving) time 
the particles spend in the active cylinder.
In our examples the Compton losses dominate over the synchrotron losses 
(see equation \ref{eq:cooling2} and Table 1). 
For runs 04, 14 and 15 the cutoff  is smoother because 
the cooling rate above $\gammac\sim 3\times 10^3$ is affected by the
Klein-Nishina  regime, making it slower.

The spectrum above the cutoff is the standard cooling spectrum $\d N/\d\gamma \propto \gamma^{-2}$,  and 
it changes to a   steeper slope at higher energies. 
This is the result of the pair cascade developing in the jet.
The cascade pairs affect  the energy distribution  
making it steeper  $\d N/\d\gamma \propto \gamma^{-3}$ \citep[e.g.][]{sve87}
above the break energy $\gamma_{\rm b}\sim 10^4$--10$^5$. Pairs of lower 
energies produce  photons that can freely escape from the jet and the 
cascade stops. 
We may note here that most of the pairs in the jet are produced by photons 
emitted in the jet. Such process does not contribute to the energy release.  
However, during the cascade the typical photon energy decreases; it leads to the 
increase of the probability of the photon escape to the external environment. 
This in turn increases the efficiency of the jet energy dissipation.

%%%%%%%%%%%%%%%%%%%%%
\begin{figure}
\centerline{\epsfig{file= 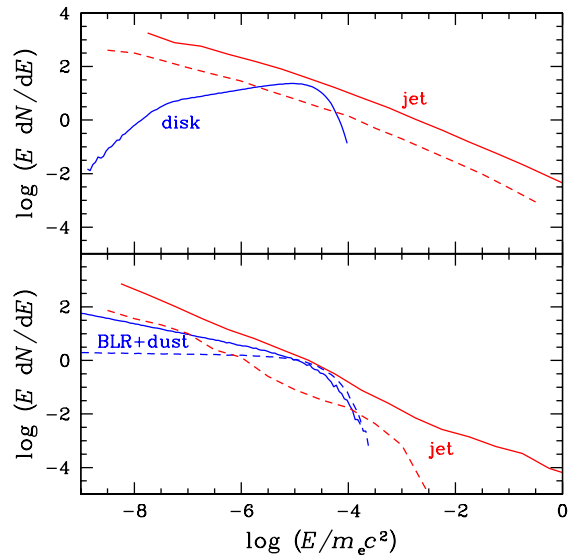,width=7.5cm}}
%\begin{center} \leavevmode \epsfxsize=7.0cm \epsfbox{nonlin.eps} \hspace{1cm}
%\epsfxsize=7.0cm \epsfbox{nonlin1.eps}
%\end{center}
\caption{The spectra of the external radiation and of the soft (mostly synchrotron) 
radiation induced by the jet activity for run 31 (solid curves) and run 35 (dashed curves). 
Notice that the spectra here are  $E\ \d N/\d E$, not  $E^2\ \d N/\d E$, to emphasise     
the number density of target soft photons available for pair production.    
Upper panel: the collimated components, 
lower panel the isotropic external radiation
and the large-angle  ($\theta>40\degr$) synchrotron jet component.}
\label{fig:nonlin}
\end{figure}
%%%%%%%%%%%%%%%%%%%%%

Fig.~\ref{fig:elz} shows the fraction of the jet energy carried by generated pairs, which 
reaches 2--5 per cent. 
The number ratio of generated electrons and positrons to that of the protons
increases with the jet power and for runs 14--10 
it is $\tau_{\pm}/\taut = 0.017, 0.135, 0.63, 1.9, 4.0$, respectively.

%%%%%%%%%%%%%%%%%%%%%%%%%%%%%%%%%%%%%%%%%%%%%%%%%%%%% 
\section{Properties and limitations of the model}
\label{sec:proper}

\subsection{Non-linear effects }
\label{sec:nonlin}

There are two kinds of non-linearity in the problem: one appears because of the 
feedback of radiation on the fluid, another is the result of the synchrotron emission, which produces 
additional opacity for the high-energy photons. 
Here we consider the second one. 
The energy density of soft synchrotron radiation of pairs produced in the outer 
decelerated layers of the  jet can exceed that of the
external radiation described in Section \ref{sec:softback}. 
In Fig.~\ref{fig:nonlin} we compare the primary external radiation to the 
jet emission averaged over the volume. 
For run 31  the latter significantly exceeds the external radiation in 
both components: longitudinal and isotropic. In this case 
the synchrotron emission of the jet becomes the main target for high-energy photons.
For runs 01, 02, 10--12 and 35, the soft radiation associated with the jet is comparable with 
the original external radiation. In runs 03, 13--15 (also runs 41 and 42 for model B), 
despite  $\etaj$  being the same,
the jet emission at low energies is much smaller than the external radiation 
because the jet spectrum is harder. 
These cases can be treated as linear, so that the solution can be scaled to any lower jet power.

In the case of substantial non-linearity, the process becomes self-supporting. The
external radiation becomes unnecessary, because the stronger synchrotron radiation 
plays the same role. \citet{sk07} have studied this kind of non-linearity  using kinetic equations
and found that a source of high-energy photons quenches itself at  high compactness.
In our case, we see that the efficiency decreases  at high compactness and the 
final spectrum gets softer. We can expect that such a tendency will continue with the
increasing compactness.

The synchrotron emission from the jet scattered in the BLR 
(neglected in our simulations) can also affect the development of the cascade  \citep{gm96}. 
The dominant contribution to the scattering is provided by the region of thickness $\sim\Rj$
immediately adjacent to the jet. 
If we associate the fraction of  radiation scattered/reprocessed in BLR, $\etaiso$, 
with the optical depth through the region of size $R$, then the optical depth 
through the layer of thickness $\Rj$ for synchrotron photons moving at angle $1/\Gammaj$ 
is of the order $\tau_{\rm sc}$$\sim$$\etaiso \theta \Gammaj$. 
We can write now the scattered luminosity~as 
\be 
L_{\rm s,BLR} = L_{\rm s}  \tau_{\rm sc} = \etas (\etaj\Ld) \etaiso\theta\Gammaj,
\ee
where $\etas=L_{\rm s}/\Lj \lesssim 0.1$ is the fraction of the jet power emitted in the 
form of synchrotron radiation. Because the scattered synchrotron is escaping 
through the surface of $\sim 4\pi \theta R^2$, the ratio of its energy density to that 
of  the BLR photons is   
\be 
\frac{U_{\rm s,BLR}}{U_{\rm BLR}} \approx \etas\etaj \Gammaj . 
\ee
This shows that the scattered synchrotron 
can be important for large $\etaj$ or $\Gammaj$.

\subsection{Temporal variability}
\label{sec:variability}

In our simulations, we have assumed that the jet power and the
Lorentz factor at the inlet do not change. However, because of the large 
non-linearity, the output photon flux demonstrates large non-Poisson fluctuations at different time-scales.
At the present level of numerical resolution, it is difficult to
distinguish confidently between numerical fluctuations and natural chaotic behaviour of the 
non-linear dynamical system.
We just would like to emphasise that the latter possibility is very probable.  
 
The transversal gradient of the Lorentz factor, $\d\Gamma/\d r$, affects the rate of the photon breeding in 
the same way as the slope of a sand pile affects sand avalanches.
The emitted radiation feeds back on the jet dynamics, making the
gradient smoother. Thus $\d\Gamma/\d r$ balances near its critical value in the same way 
as the slope of the sand pile does, when one pours sand on the top.
It seems that we observe this in the cases of highest non-linearity, when the 
soft radiation of the jet dominates the external soft radiation (runs 10 and 31, see Fig.~\ref{fig:nonlin}). 
It these cases the radiation power is more sensitive to the transversal 
gradient of $\Gamma$ because of a short free path of high-energy photons in this direction. 

If we assume a large fluctuation in the jet power at a time-scale shorter than $R/c\Gammaj^2$, 
the photon avalanche will be propagating along the jet and the  
observer will see a large increase in the luminosity partially  because of the time compression. 
Simulations of strongly variable sources in the context of photon breeding model 
are left for future studies (see also SP06).

\subsection{Impact on the external environment}
\label{sec:external}

The high-energy photons produced in the jet interacting outside of the jet 
deposit their momentum into the external environment and therefore accelerate it. 
This acceleration can be neglected if the density of the external matter is sufficiently high. 
Let the momentum deposited in a unit volume per unit time is
$\kappa(r,z) = \d p/(\d V \d t)$. Then, equating the work of the photon pressure force 
to the kinetic energy of the flow one gets the final velocity of 
the entrained medium (in non-relativistic limit):
\be
\beta(r) = v/c = \left( \frac{2 \int \kappa(r,z) \d R}{ n \mpr c^2} \right)^{1/2}, 
\ee 
 where $n$ cm$^{-3}$ is the particle density of the external medium. 
The required condition $\beta \ll 1$, allows to put the lower limit 
of the density of the environment: 
\be \label{eq:ncrit}
n \gg n_{\rm cr} \equiv \frac{2  \int \kappa(r,z) \d R }{\mpr c^2} . 
\ee 

During the simulations we accumulate the average deposited momentum  $\kappa$. 
Its dependence on $r$ is strong. 
For  the layer closest to the jet, $1 < r < 1.005$, we have $\int \kappa (1<r< 1.005,z) \d R = 17$ erg cm$^{-3}$ 
for run 31,  which translates to $n_{\rm cr}=2\times 10^4$ cm$^{-3}$. 
The requirement (\ref{eq:ncrit}) seems reasonable for the BLR, where 
the cloud density is about $10^9$ cm$^{-3}$ \citep{of06} and even the 
mean density should not be much smaller.
 The critical densities for other runs are even smaller: 
$n_{\rm cr}=10^3$ cm$^{-3}$  for runs 02 and 12, 
$n_{\rm cr}=30$ cm$^{-3}$  for runs 04 and 14.
For runs 41 and 42, the required density is $n_{\rm cr}=10$ cm$^{-3}$,  
which is still reasonable at a parsec scale.

\subsection{Important phenomena beyond ballistic approximation}
 \label{sec:ballis}
 
Our consideration is exact and model independent when concerning photons and pairs: 
their interactions, propagation, breeding, etc. at a given distribution 
of the fluid Lorentz factor and a given  magnetic field. 
The only assumption we make is the spectrum and density 
of seed high-energy photons, this is however not important as their 
energy content is negligible and the system ``forgets'' 
their spectrum very soon. 

On the other hand, our consideration is very approximate and is based on a 
simplified model when it concerns the fluid and the feedback of the particles 
on the fluid. How our simplifications affect the results and what we would 
obtain with a more realistic hydrodynamical treatment of the problem?

First, the pressure will redistribute the Lorentz factor of the fluid.
This will hardly change significantly the process of particle breeding.
More important,  the formation of internal shocks at regions of maximal deceleration
can increase the radiation efficiency by accelerating charged particles.

Then, we can expect that a turbulent layer near the jet boundary 
will be formed due to the large shear gradient of the Lorentz factor.
The same can happen due to the Kelvin-Helmholtz instability of the 
boundary discontinuity \citep{dy86}, however for an ultra-relativistic jet 
this instability develops slowly. The radiative deceleration of  the outer 
layer of the jet can generate the turbulent mixing layer  much earlier.
Then the  turbulent layer should again contribute to the radiation 
due to diffusive particle acceleration \citep[see][]{so03}. 

Finally, we ignore adiabatic heating of relativistic electrons while the jet 
(comoving) compression ratio in $z$ direction is $\sim \Gammaj/\Gamma$, 
where $\Gamma$ is the Lorentz factor of the decelerated fluid. The account of
the adiabatic heating is not simple, because the jet can expand in the transversal direction. 
The compression also leads to the amplification of the magnetic field $\Bj\propto 1/\Gamma$ 
(ignored in our simulations), which would in turn increase the role of synchrotron radiation in the 
decelerated part of the jet. 

The formation of shocks,  turbulence and adiabatic heating of electrons, 
which  can be caused by the jet deceleration,  
should  increase the radiative efficiency of photon breeding mechanism,
while the magnetic field amplification probably decreases it.
These phenomena are interesting  and should be studied in  future simulations.

\subsection{Origin of the seed high-energy photons}
\label{sec:highseed}

In the presented simulations  the energy content of the seed high-energy photons, 
which initiate the further evolution of the system, was taken to be 5--7 orders of 
magnitude smaller that the final steady-state total energy content in photons within the jet.  
Where these initial photons can originate from? They can be 
associated with the shock acceleration of charged particles in the jet. 
There is, however, a  minimum possible source of high-energy, seed photons 
provided  by the extragalactic gamma-ray background. 
The observed gamma-ray background intensity at Earth at $\phe \gtrsim 10^5$  is  
$\sim  10^{-6}$ GeV cm$^{-2}$ s$^{-1}$ sr$^{-1}$ \citep{sree98}, corresponding 
to the energy density $6\times 10^{-19}$ erg cm$^{-3}$. This 
would give the energy $\sim  3\times 10^{31} R_{\rm j,16}^3$ erg 
within the considered cylindrical volume of radius $\Rj$ and height 20$\Rj$.

%%%%%%%%%%%%%%%%%%%%%%%%%%%%%%%%%%%%%%%%
\begin{figure}
\centerline{\epsfig{file=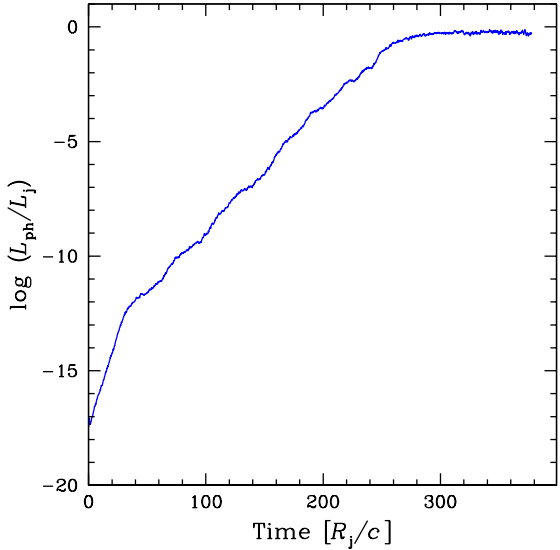,width=7.5cm}}
\caption{Fraction of the jet power converted into photons versus time for 
run 30. The initial density of the high-energy, seed photons corresponds to the 
extragalactic gamma-ray background.
}
\label{fig:lowstart}
\end{figure}
%%%%%%%%%%%%%%%%%%%%%%%%%%%%%%%%%%%%%%%%

We thus made a simulation (run 30) where the total energy of 
the seed high-energy photons in the simulation volume was 20 orders of magnitude smaller 
(approximately corresponding  to the energy of the extragalactic gamma-ray background)
than the jet energy in the same volume $E_{\rm j}\sim 10^{51}$ erg. 
The result is shown in Fig.~\ref{fig:lowstart}. The system passes 20 orders of 
magnitude in the total photon energy  during 300$\Rj/c \sim$3 years. 
A jump  at the very beginning of simulations corresponds to almost immediate conversion of  
seed photons into pairs 
with the corresponding energy increase by a factor $\Gammaj^2=400$. 
The photon energy grows exponentially with the time constant $\taupb = 2.7$ at $t \lesssim 25$ 
and $\taupb = 8.2$ at later time. 
The   rapid rise at $t <20$ corresponds to the growth of the photon avalanche 
as it moves downstream with the flow.  At later times the growth is supported by the spatial feedback loop
due to up-streaming photons.  
The final steady-state regime does not depend on the initial density of high-energy photons.
This numerical experiment means that a jet, 
with parameters corresponding to run 30, would radiate half of its total energy in 
three years just due to the photon breeding starting  off the extragalactic gamma-ray background.

%%%%%%%%%%%%%%%%%%%%%%%%%%%%%%%%%%%%%%%%%%%%%%%%%%%%%
\section{Astrophysical implications}
\label{sec:astro}

\subsection{Spectral energy distribution of blazars}
 \label{sec:sychmax}
 
Blazars demonstrate two-component spectra.  The low-energy 
component peaking in the IR--X-ray range (the flares in
the TeV blazars show hard X-ray spectra peaking above 100 keV) is 
traditionally interpreted as synchrotron radiation from relativistic electrons. 
The high-energy component   is the Comptonized radiation from some seed soft photons. 
The spectra of low-luminosity objects (TeV blazars, BL Lacs) are well described by the 
synchrotron self-Compton (SSC) models. 
In high-luminosity blazars, both synchrotron  and Compton components peak at lower energies
(so called blazar sequence, \citealt{fmc98};  but see \citealt{gio07}).
The role of the external radiation (from BLR or dusty torus) 
in providing seed photons is probably more important, and 
the spectra can be described by external radiation Compton (ERC) model  \citep{s94,sik97}. 
In all objects, the two components are rather well separated.  

The spectra, predicted by  the photon breeding model (see Fig.~\ref{fig:spec}a), 
demonstrate similar features. 
At low luminosities and compactnesses (runs 04, 14-15),
 the mean Lorentz factor of the electron-positron distribution is large (see Fig.~\ref{fig:espec}).
In spite of the fact that (in the jet frame) the magnetic energy density  is 1--2 orders of magnitude 
smaller than that of the isotropic radiation (i.e.  $\ellb\ll \elli\Gammaj^2$, see Table 1), the 
synchrotron power strongly dominates over the external Compton, because the external 
photons interact with the high-energy pairs in the Klein-Nishina regime. 
As most of the synchrotron photons are also above the Klein-Nishina cutoff $\sim 1/\gamma$, 
the SSC is weaker than the synchrotron. 

At higher luminosities (and compactnesses), the high-energy photons produce a   
pair cascade. This reduces the mean energy per pair (see Fig.~\ref{fig:espec}, runs 01--03, 10--13). 
At these compactnesses the pairs also cool more efficiently producing a cooling distribution 
$\d N/\d \gamma\propto \gamma^{-2}$  (see Section \ref{sec:pairs}), which changes 
to a steeper $\gamma^{-3}$ behaviour at higher energies as a result of pair cascade. 
The latter part of the pair energy distribution is responsible for the flat part of 
photon spectrum (in $E^2 \d N/\d E$ units).\footnote{One should note here that the pair cascade does 
not degrade the pair energies  infinitely, when the energy of  the photon in the jet frame drops below 
$1/\Thetamax\Gammaj\sim10^{4}$, it can easily escape and the cascade stops. }
The reduction of the mean pair energy results in a shift of 
the photon spectra to lower energies (Fig.~\ref{fig:spec}a). 
It also leads to the increased role of the ERC, because now most of external photons interact 
with pairs in Thomson regime.  

The properties of our model are the increasing role of the ERC over SSC and synchrotron 
with luminosity and a shift of the spectral peaks to lower energies. These are in general agreement with 
the observational trends. 
However, there are some discrepancies between the shapes of the 
theoretical and the observed spectral shapes. 
At low luminosities, the  synchrotron peak in our model is in the 100 MeV--1 GeV range, while it is 
normally observed in the X-ray range. 
One need to note here that the angle-dependent spectra at $\theta\gtrsim 1/\Gammaj$ 
have sharper features (see Fig.~\ref{fig:spe_ang}) 
and the synchrotron spectrum peaks at slightly lower energies that the spectrum averaged 
over $10\degr$ around the jet axis. 
Another possible reason for the discrepancy is that the electron distribution is broad 
 and extends to very high energies. 
This is a consequence of our assumed spectrum of the soft isotropic radiation 
extending to a very low energy $x_{\min}=10^{-9}$. 
%Taking $x_{\min}=10^{-7}$ (which corresponds to the temperature of $600$ K), would significantly 
%reduce the range of  photon energies, which can interact with soft photons 
%within the jet, producing pairs, would shift the synchrotron peak to lower energies and make 
%a separation between the spectral components more pronounced. 
A narrower and harder distribution of external soft photons  might make the injection 
function of pairs narrower and softer, so that the maximum pair energy is below $\gamma_{\max}$
given by equation (\ref{eq:gmax}), reducing thus the energy of the synchrotron cutoff.

At high luminosities, our model predict basically a single component where  the  synchrotron and ERC components 
are mixed, while there are clearly two separated components in the data. 
However, there is no consensus whether the two components are co-spatial.  For example, 
the bright blazar 3C279  did not show almost any variability in the radio--optical band, 
when the gamma-ray  flux changed by an order of magnitude \citep{weh98}.  
Thus the radio-to-optical emission in the high-luminosity sources can be produced is a separate region,   
far  away from the black hole. 

In our model, the pairs escaping  from the region filled with soft photons  
still carry a considerable (a few per cent) fraction of the jet energy (see Fig.~\ref{fig:elz}).   
If these pair are reheated at larger scales (e.g. by a diffusive acceleration by various plasma perturbations),  
where soft radiation field is weaker, they will radiate most of their energy in the form of synchrotron.
At the expense of 3--5 per cent of the jet energy for pair reheating, 
the synchrotron peak observed in low-peak blazars could be reproduced. 
If this hypothesis is valid then the high- and the low-energy peaks
have different origin: the high-energy peak is a mixture of prompt  synchrotron 
and Compton radiation of high-energy pairs in the fast  cooling regime.
The low-energy peak is the synchrotron radiation of pairs in a heating/cooling balance. 
In such a scenario the intensity of high- and low-energy peaks should correlate 
in time, but in a loose manner, the correlation should be stronger at longer time-scales.

\subsection{Electron energy distribution in BL Lac objects and particle acceleration}

One of the most often used in the recent years scenario for formation of the relativistic electrons, 
responsible for the observed emission, involves the diffusive (Fermi) acceleration at relativistic 
collisionless shocks (or shear layers). 
It has been claimed that the acceleration  at relativistic shocks produces the power-law 
particle spectrum $\d N/\d \gamma\propto \gamma^{-p}$ with 
a universal $p=2.2$ index \citep[see e.g.][]{agk01,kw05} which is close 
to the value observed in BL Lacs \citep*{gcc02}. 
These results, however, were based on certain simplifying assumption regarding the 
scattering of particles by magnetic inhomogeneities. 
Using  accurate Monte Carlo simulations with a more realistic magnetic field structure,
\citet{no06} demonstrated that no universal power-law is produced 
and questioned the role of the Fermi-type particle acceleration mechanism in relativistic shocks.
%In addition, diffusive acceleration of electrons (and positrons) requires that 
%they are heated/pre-accelerated to sufficiently high energies $\gamma\gtrsim \mpr/\me$
%before they even can cross the shock front which thickness is defined by the Larmor 
%radius of thermal protons. 
 
Using particle-in-cell simulations,  \citet{spi07} recently showed that the electrons 
can be heated by interaction with ion current filaments in the upstream region 
reaching almost an energy equipartition with protons.  The low-energy cutoff appears at
$\gamma_{\min}\sim \frac{1}{5}\Gamma \mpr/\me$.  
In internal shock model, the relative velocity of colliding shells is 
mildly relativistic and thus we expect $\gamma_{\min}\sim500$.
At this stage,  it is too early to say whether the model would be able to produce a power-law 
electron distribution extending to very high energies.

The detailed fitting of the broad-band spectra of TeV blazars and low-power BL Lacs 
with the SSC model shows that the distribution of relativistic electrons injected to the system can be 
modelled as a power-law of index $p\approx 2$--2.5 with the
low-energy cutoff at $\gmin \sim 10^4$--$10^5$ and 
the high-energy cutoff at $\gmax\sim 10^6$--$10^7$ 
\citep{gcc02, kca02,kon03,gdk07}.\footnote{The high-luminosity objects require  
much lower energy electrons \citep{gcc02}, but this result may be biased because 
the electron distribution is obtained from one-zone models, while 
the synchrotron peak in reality may be not related to the gamma-rays.}
Such a peaked distribution of injected electrons is not consistent with a diffusive process. 

The photon breeding mechanism, on the other hand, predicts the injection spectrum 
to be bounded and to mirror (relative to $\me c^2$) 
the spectrum of the soft photon background, because the pairs in the jet are produced only
by those high-energy photons that can interact with the soft photons. 
For example, if soft photons are from the accretion disc of maximum temperature $\Thetamax$, 
the low-energy injection cutoff is at $\gmin \sim \Gammaj/6\Thetamax$ 
(see SP06; factor $\Gammaj$ comes from the Lorentz transformation to the jet frame). 
This cutoff should depend weakly  on the luminosity and the black hole mass, 
which defines the characteristic emission radius, so that 
$\Thetamax\approx 10^{-5}\Ldfv^{1/4}(M_{\rm BH}/10^8\msun)^{-1/2}$ and thus 
we can predict that 
\be \label{eq:gmin}
\gmin \sim 3\times 10^5 \left(\frac{\Gammaj}{20}\right) \Ldfv^{-1/4} 
\left( \frac{M_{\rm BH}}{10^8\msun}\right) ^{1/2}. 
\ee
At high compactnesses, however, the synchrotron photons from the jet 
can provide enough opacity (see  Section \ref{sec:nonlin}) and $\gmin$ may be smaller. 

For the conditions appropriate for AGNs, the high-energy cutoff of the pair distribution $\gmax$, 
that can participate in photon breeding, is defined by the different physics. 
An electron (or positron) of Lorentz factor higher than that given  by equation (\ref{eq:gmax})
loses most of its energy to the synchrotron emission which escapes freely from the jet. 
Thus, at low compactnesses the pairs are injected to the jet between  $\gmin$ and $\gmax$. 
At high compactnesses, because of the higher opacity the Compton scatted photons cannot easily 
escape from the jet, producing a pair cascade which degrade their energy 
(see Fig.~\ref{fig:espec}). 
The range of Lorentz factors of injected pairs as predicted by the photon breeding mechanism 
is in general agreement with observations of BL Lacs.

 \subsection{Jet power and composition}

In many models of the high-energy emission from relativistic jets, 
the radiative efficiency is low because only  internal energy of the jet can be  dissipated  
(e.g. in internal shock model) 
and only a small fraction of the energy is transferred to the (high-energy) electrons. 
This leads to the  large jet power requirements. 

The photon breeding mechanism has important properties that significantly 
reduce the energy budget required: (i) it produces only electrons (positrons) of high Lorentz factors;
(ii) it efficiently decelerates the jet, using thus most of its total power. 
Thus the radiative  efficiency is extremely high, reaching in many cases 50--80 per cent.

The primary composition of the jet is not constrained if the photon breeding works, 
except the general requirement that it should not be dominated by cool pairs. 
Indeed, a dynamically important component of cold pairs can be excluded because of  the absence
of the bulk Compton scattering bump produced by such pairs in the soft X-rays \citep{sm00}.
Moreover, large contribution of pairs to the bulk energy might be problematic, because 
of their short annihilation time in the vicinity of the black hole  \citep{bl95}. 
Magnetically dominated jet is less favourable  for photon breeding at $\Gamma < 20$, 
while for a larger Lorentz factor the radiative efficiency is not sensitive to the jet magnetisation. 

Independently of the primary jet composition the photon breeding loads the jet
with a significant amount of relativistic pairs. The contribution of these pairs to the jet energy 
flux according to our simulations is 2--5  per cent (see Section \ref{sec:pairs}). 
This is probably an underestimation as we do not take into account  diffusive acceleration 
and adiabatic heating of pairs. 
The number ratio of generated electrons and positrons to protons (in the case of 
matter-dominated jet) may be related to the mean electron random Lorentz factor 
(as measured in the jet frame) and the radiative efficiency: 
\be
\frac{\tau_{\pm}}{\taut } \approx \frac{\mpr }{\me}  \frac{\etaeff} {\langle \gamma \rangle } .
 \ee
For low-power jets, a typical $\gamma\sim10^5$--$10^6$, and 
the number of generated pairs should be relatively small. 
At high powers, the mean pair energy decreases and 
the generated pairs   dominate (Section \ref{sec:pairs}). 
Thus, for high-power quasars the model predicts the number of pairs 
exceeding in number the protons, while dynamically the protons still dominate. 
This is consistent with the constraints on the pair content obtained 
from the X-ray spectra of blazars associated with quasars  \citep{sm00}.

\subsection{Conditions for photon breeding and gamma-ray emission sites}

The arguments based on the gamma-ray transparency and the observed short time-scale variability  
constrain the gamma-ray emitting region in blazars to distances about $\sim10^{17}$ cm \citep{s94}.
This distance is  often associated with the BLR. 
In the internal shocks models, the shells of  Lorentz factors $\Gamma_1$ and 
$\Gamma_2$ ejected from the central source and initially separated by the distance $R_0$  collide 
at the distance $R=R_0 \Gamma_1^2/[1-(\Gamma_1/\Gamma_2)^2]$. Taking 
$R_0=10^{15}$ cm ($\approx 30 \Rg$ for a $10^8\msun$ black hole), 
$\Gamma_1=10$ and $\Gamma_2\gg \Gamma_1$ we get $R=10^{17}$ cm. 
However, the agreement with the aforementioned constraints is rather coincidental, because 
the black hole mass can vary by 3--4 orders of magnitude and there is no physical 
theory yet predicting the assumed separation between the shells. 

The photon breeding mechanism allows various emission sites. 
If the jet is already accelerated and collimated at distances $\sim100 \Rg$ from the black hole, 
the direct radiation from the accretion disc can serve as a target for high-energy photons 
for pair production and to trigger the photon avalanche.
The photon-photon opacity through the jet is very large at these distances and the 
jet deceleration and radiative efficiency might be rather small, because only a very thin boundary layer is active. 
The escaping radiation  should have a high-energy cutoff at tens of GeV rather than at TeV energies. 
At larger distances the disc radiation becomes more beamed along the jet and the disc photons 
do not interact anymore with the high-energy radiation produced in the jet. Thus the photon breeding stops. 
The jet now is, however, loaded with high-energy photons that can serve as seed high-energy radiation for the 
avalanche to be triggered at larger distances, where the conditions for photon breeding are satisfied. 
The natural scale for that is the position of the BLR, where the density of the isotropic soft photons 
is sufficiently high. As we showed in Section \ref{sec:results}, the radiative efficiency here can be 
very large.  
 
At even larger distances, the BLR photon density drops, the jet becomes transparent for high-energy 
photons and the photon breading stops.
If at the parsec scale, the jet is still relativistic enough,   
the process can start operating again on the IR radiation from the dusty torus.   
The typical dust temperature is  
$\Theta_{\rm dust}=kT_{\rm dust}/\me c^2 \approx 10^{-7} \Ldfv^{1/4} R_{\rm pc}^{-1/2}$. 
The pair-production optical depth is about  
\beq \label{eq_tauggdust}
\taugg & = & n_{\rm ph,dust} (0.2 \sigmat) \Rj 
 =  30 \etaiso (20\theta) \frac{\Ldfv}{\Rpc \Theta_{\rm dust,-7}}  \nonumber \\
 & = &  30 \etaiso (20\theta) \Ldfv^{3/4} \Rpc^{-1/2}   ,
\eeq
where $n_{\rm ph,dust}=0.37\ U_{\rm dust}/kT_{\rm dust}$ is the number density of photons from the 
dust and $\etaiso$ is the ratio of the dust to the disc luminosities.  
Thus, even for $\etaiso$ of the order of a few percent, the photon breeding
is effective at the parsec scale in bright quasars. 

It has not escaped our attention that the jet may become active  even further out: 
by interacting at kpc scale with the stellar radiation field 
and  at $\sim$100 kpc with the cosmic microwave background radiation. 
 
We note here that  photon breeding works well at high disc luminosities (see Section \ref{sec:radeff})
characteristic for quasars. In BL Lac objects, where the discs are less luminous, 
photon breeding may operate at small distance scale, which in turn requires 
small black hole mass and/or fast jet acceleration, or for  large jet Lorentz factor $\Gammaj\gtrsim 40$. 
The observed fast variability from PKS 2155--304 \citep{aha07} and 
Mrk 501 \citep{alb07} supports these suggestions.

\subsection{The Doppler factor crisis in TeV blazars}

The Doppler factors required by the homogeneous SSC models to describe the spectra of the 
blazars emitting at TeV energies (to avoid absorption by the IR radiation), 
is rather large $\doppler$$\sim$20--50  \citep{gcc02, kca02,kon03,gdk07}.  
Such high $\doppler$ values disagree with the small apparent velocities observed at the parsec scale 
in  Mrk 421 and Mrk 501 \citep{mar99,pe05} and other TeV blazars \citep{pe04}.  

%\subsubsection{Jet deceleration}

One proposed solution to this $\doppler$-crisis \citep{tav05} is that the jet Lorentz factor   
drops from 15 at the gamma-ray emitting, subparsec scale to $\sim$4 at 
the radio-emitting VLBI parsec scale  \citep{gk03}. 
For the viewing angle within the beaming pattern of the initial jet, $\theta\lesssim 1/\Gammaj$, 
the apparent velocity of the decelerated jet with Lorentz factor $\Gammaf\ll\Gammaj$  is 
\be
\beta_{\rm f, app} = \frac{\beta_{\rm f}\sin\theta}{1-\beta_{\rm f}\cos\theta}
 \approx \frac{2\theta\Gammaf^2}{1+(\theta\Gammaf)^2}  
\approx 2 \theta \Gammaf^2 \lesssim 2 \frac{\Gammaf^2}{\Gammaj} . 
\ee
Thus, deceleration of the jet to $\Gammaf< \sqrt{\Gammaj/2}$ guaranties 
that the apparent motion is subluminal.

The question arises what is the mechanism that decelerates the jet  at subparsec scale. 
The internal shocks in the jet \citep{spa01}   are rather  
inefficient unless huge fluctuations of the Lorentz factors are involved 
\citep{amb00}, which seems unlikely. 
On the other hand, the photon breeding mechanism predict significant  deceleration of the jet 
(see Table 1) and strong emission at subparsec scale, if the conditions for photon breeding are satisfied. 
The photon breeding thus may be responsible for jet deceleration and resolution of the 
$\doppler$-crisis.
 
A different way to solve the $\doppler$-crisis is to assume that the jet consists of  
a fast spine and a slow sheath \citep{ccc00,gtc05}. In this model a slower sheath 
having a softer spectrum dominates the emission in the radio band. 
The photon breeding mechanism produces exactly  such a   structure (Fig.~\ref{fig:lorlev}) and 
the spectra (Fig.~\ref{fig:angle}) in a self-consistent way.

Thus, the jet behaviour  predicted by the photon breeding mechanism unites
the two proposals: it decelerates the jet at subparsec scales and produces a 
fast spine--slow sheath structure. 
 
\subsection{Emission pattern}

\subsubsection{Unification of BL Lacs and FR I radio galaxies}

According to the unification scheme of radio-loud active galaxies \citep{up95}, 
BL Lac objects are Fanaroff-Riley I (FR I) radio galaxies with their jets oriented close to the line of sight. 
The average Lorentz factor of the jets, derived by matching the luminosity functions 
and statistics of BL Lac and FR I (and B2) samples, was estimated to be $\Gammaj \sim$3--5 \citep{up95,har03}. 
The  core optical  luminosities of  BL Lacs are on average $10^{4.5}$ times higher than those of FR I \citep{ccc00} 
with the same extended radio luminosity.
Associating the core emission with the jet, we again get rather low Lorentz factors  $\Gammaj=6$ or 4 for the 
continuous steady jet  or a moving blob, respectively. 
These low $\Gammaj$ are in clear disagreement with the much higher values of the Doppler factors required 
 to explain the gamma-ray emission of BL Lacs. 

Both decelerating \citep{gk03} and structured \citep{ccc00,gtc05} jet  models provide explanations for the discrepancy. 
Because the  photon breading mechanism produces decelerating and structured jet, it obviously is also 
capable of unifying BL Lacs with the radio galaxies, keeping high $\Gammaj$ to explain the gamma-ray emission. 
In blazars the emission from the fast, $\Gammaj\sim20$, spine dominates.  
In radio galaxies observed at large angles $\theta\sim1$ (see Fig.~\ref{fig:angle}), 
the optical emission  is produced in the slower sheath  of $\Gamma=$2--6.

\subsubsection{Off-axis emission and the TeV emission from radio galaxies}

As we have discussed above, the photon breeding mechanism naturally produces 
very broad beams of photons \citep*[see also][]{der03,der07}. 
The gamma-ray emission at $\theta\sim1$ is produced in the external environment 
(see Fig.~\ref{fig:angle}).  The luminosity ratio between the nearly isotropic emission
and the beamed emission at $\theta\approx1/\Gammaj$    is about $\Gammaj^{-4}$  
(two powers of $\Gammaj$ come from the energy amplification 
in the jet and two powers appear because of beaming). 
This ratio is $\sim\Gammaj^2/10$ larger than that predicted by equation (\ref{eq:isojet}). 
Thanks to the emission from the external medium, the nearby misaligned  jets 
become observable in high-energy gamma-rays. 

The central source of M87, which has the best studied jet, has 
a rather low disc luminosity of only  $\sim10^{42} \ergs$ \citep*{bsh91}.
With the threshold $\Ld >10^{43} R_{17} \ergs$ for $\Gamma=20$ 
the photon breeding in model A  is possible at $R < 10^{16}$ cm, 
which is only $\sim 10 \Rg$ for the estimated 
mass of the central black hole $\sim 3 \times 10^9 \msun$   \citep{mac97}. 
At such a distance the photon breeding will operate on the direct disc radiation (model C). 
The implied distance is consistent with the detection of the 
rapid  ($\sim10^5$ s) variability of the TeV photon flux from M87 \citep{aha06}.
However, photon breeding may not operate, if the wide pattern at the base of the jet 
observed by VLBI (\citealt{jun99}, but see \citealt{kri07}) 
actually reflects the jet geometry, which would mean that the opening angle is large. 
 
\subsection{Observational appearance and volume dissipation}
 
Most of the emission from the jet undergoing photon breeding comes 
from the regions of maximum gradient of $\Gamma$. In the steady-state, 
the volume emissivity contributing to the observed emission (at small angles to the jet) is 
 \be
j(r,z) \propto \frac{\d \Gamma(r,z)}{\d z} \frac{\doppler^3}{\Gamma} . 
\ee
In the case of small luminosities (and small $\Gammaj$) only the outer layers of the jet
suffer significant deceleration (see Fig.~\ref{fig:lorlev}b, run 24), resulting in the limb-brightening,
with most of the emission coming from large $z$, where the cascade had time to develop. 
The situation is more complicated  at high luminosities (or large  $\Gammaj$). 
The outer layers decelerate rapidly (see curves for runs 21--23 in Fig.~\ref{fig:gamz}) 
making the emission limb-brightened at small $z$, while the layers close to the jet axis decelerate slower. 
Thus the emission from the core dominates  at large $z$. 
At angles larger than $\gtrsim 1/\Gammaj$, only slower, significantly  decelerated layers contribute 
to the observed luminosity, and therefore we always would see the limb-brightened emission. 
 
Our predictions are consistent with the observed limb-brightened  morphology of the Mrk 501
jet at the parsec scale  \citep{gir04}. 
Such a  structure  could  a result from the efficient deceleration and  loading by relativistic pairs 
of the jet outer layers by the photon breeding  at a subparsec scale.

\section{Conclusions}

\label{sec:concl}

The photon breeding mechanism can be an extremely efficient way 
of dissipating bulk energy of relativistic ($\Gammaj\gtrsim 10$) jet into high-energy radiation. 
In the case of high luminosity AGNs and powerful jets, the mechanism is 
very robust: it works at any value of the jet magnetic field and is not
sensitive to the spectrum of the external radiation. At high luminosities 
and jet Lorentz factors, the mechanism should work almost inevitably, the only 
fact that can prevent the runaway photon breeding is a 
purely longitudinal  (along the jet) geometry of the magnetic field.

At the intermediate range of  AGN luminosity, $10^{43} <  \Ld R_{17} < 
3\times 10^{44} \ergs$ and the Lorentz factor above 10, the mechanism 
still can work efficiently, but only at favourable conditions: 
a soft spectrum of external isotropic radiation and a weak (dynamically not 
important) magnetic field. 
When the Lorentz factor is below $\sim 8$ or the AGN luminosity is below
$10^{43} R_{17}\ergs$ the photon breeding mechanism does not work.

The mechanism can work in different astrophysical contexts: in the immediate vicinity of  
the accretion disc,  in the broad line region (probably, most efficiently), and 
in the IR radiation field produced by the dust at a parsec scale. 
There could be other sites for operation of the photon breeding mechanism 
(e.g. stellar radiation field at a kpc scale, cosmic microwave background radiation at 100-kpc scale). 

There exist a clear signature of the mechanism which can be observed  by {\it GLAST}
in the spectra of moderately bright blazars: a synchrotron cutoff at a few GeV range.
 
There exist some problem with the straightforward interpretation of the blazar spectra.  
The low-energy synchrotron peak reproduced by our simulations is less prominent than 
the one observed in the spectra of gamma-ray quasars.
An additional mechanism of gradual pair reheating is necessary. 
This could be associated with a diffusive Fermi acceleration of pairs, produced 
by the photon breeding, by fluid perturbations. Such reheating mechanism should have a moderate 
energy budget and seems very natural after a strong impact of radiation on the jet.
Importantly,  the photon breeding solves the problem of injection of relativistic electrons 
for Fermi acceleration: they are already produced by high-energy photons.

Fast deceleration of the outer jet layers results 
in  the fast spine--slower sheath structure which is also implied by various observations. 
Particularly, the differential deceleration can resolve the contradiction between a 
requirement of the large Doppler factor in the
gamma-ray emitting region and the low Lorentz factor of jets in
BL Lacs derived from the radio and optical observations.

Despite the fact that the photon breeding mechanism has not been known before 
and may seem somewhat exotic, it is actually much simpler in the description than the
diffusive Fermi acceleration, which depends on complicated plasma phenomena. 
Once the velocity pattern for the fluid, the magnetic field and the external radiation field are specified, 
the fate of each high-energy photon and its descendants can be reproduced (in statistical sense) from 
first principles, because the interaction cross-section are known with high accuracy. 
The question whether a photon produces a runaway avalanche or not, can be answered exactly. 
However, there still remain some details which require a more elaborated treatment of the
fluid dynamics and a more complete model for the jet. 
Particularly, it would be important to study the jet launch, the photon breeding, and the 
jet deceleration with the resulting gamma-ray emission in one numerical experiment.

%----------------------------------
\section*{Acknowledgments}
 
The work is supported by the Russian Foundation for Basic Research grant  07-02-00629-a,
the Magnus Ehrnrooth Foundation, the Vilho, Yrj\"o and Kalle V\"ais\"al\"a Foundation, and  
the Academy of Finland grants 110792 and 112982.
We thank Anatoly Spitkovsky and Amir Levinson, the referee,  for comments. 
%----------------------------------

%----------------------------------
\label{lastpage}

\end{document}